\documentclass[useAMS,usenatbib]{mn2e}
\usepackage{graphicx,amssymb,color}
\usepackage[fleqn]{amsmath}
\usepackage[normalem]{ulem}
\newcommand{\BigO}[1]{\ensuremath{\operatorname{O}\bigl(#1\bigr)}}  

\title[Liberating exomoons in white dwarf planetary systems]
{Liberating exomoons in white dwarf planetary systems}
\author[Payne, Veras, Holman, \& G\"{a}nsicke]{
Matthew J. Payne$^{1}$\thanks{E-mail:matthewjohnpayne@gmail.com, mpayne@cfa.harvard.edu}  
Dimitri Veras$^{2}$,
Matthew J. Holman$^{1}$,
Boris T. G\"{a}nsicke$^{2}$
\\
$^{1}$Harvard-Smithsonian Center for Astrophysics, 60 Garden St., MS 51, Cambridge, MA 02138, USA
\\
$^{2}$Department of Physics, University of Warwick, Coventry CV4 7AL, UK
}

\newcommand{\gsim}{\lower.7ex\hbox{$\;\stackrel{\textstyle>}{\sim}\;$}}
\newcommand{\lsim}{\lower.7ex\hbox{$\;\stackrel{\textstyle<}{\sim}\;$}}

\begin{document}
\pagerange{\pageref{firstpage}--\pageref{lastpage}} \pubyear{XXXX} 
\maketitle
\label{firstpage}

\begin{abstract}
Previous studies indicate that more than a quarter of all white dwarf (WD) atmospheres are polluted by remnant planetary material, with some WDs being observed to accrete the mass of Pluto in $10^6$ years.
The short sinking timescale for the pollutants indicate that the material must be frequently replenished. 
Moons may contribute decisively to this pollution process if they are liberated from their parent planets during the post-main-sequence evolution of the planetary systems.
Here, we demonstrate that gravitational scattering events amongst planets in WD systems easily triggers moon ejection.  
Repeated close encounters within tenths of a planetary Hill radii are highly destructive to even the most massive, close-in moons.  
Consequently, scattering increases both the frequency of perturbing agents in WD systems, as well as the available mass of polluting material in those systems, thereby enhancing opportunities for collision and fragmentation and providing more dynamical pathways for smaller bodies to reach the WD. 
Moreover, during intense scattering, planets themselves have pericenters with respect to the WD of only a fraction of an AU, causing extreme Hill-sphere contraction, and the liberation of moons into WD-grazing orbits. 
Many of our results are directly applicable to exomoons orbiting planets around main sequence stars.
\end{abstract}


\section{Introduction}\label{SECN:INTRO}

A number of observed features suggest that not only do planetary systems exist around white dwarfs (WDs), but that these systems are dynamically active.
These signatures come in three forms: 
(1) direct detection of major or minor exoplanets, 
(2) heavy metal pollution in WD atmospheres,
and 
(3) debris discs that surround WDs.  

Direct detections include the disintegrating minor planet (or planets) which has been observed orbiting WD 1145+017 with an orbital period of under 5 hours\footnote{This planet represents the smallest and quickest substellar body that has so-far been observed.} \citep{vanetal2015,2015arXiv151006434C} and one very wide orbit ($\sim 2500$ au) 
super-Jovian but sub-brown-dwarf mass companion \citep{luhetal2011}.

WD atmospheres are chemically stratified such that only the lightest elements do not sink
below the convective layer.  
The sinking timescales of the heavier elements are so quick (typically days to weeks) -- see Fig. 1 of \cite{wyaetal2014} -- that the presence of metals in the atmospheres is refereed to as ``pollution''.
Between one-quarter and one-half of all single WDs in the Milky Way are  metal-polluted \citep{zucetal2003,zucetal2010,baretal2014,koeetal2014}, a range that is commensurate with the fraction of Milky Way MS stars which are thought to host planets \citep{casetal2012}.  
The metal pollution almost certainly predominantly arises from planetary remnants,  as in WD 1145+017.
An accretion origin from the interstellar medium has been ruled out 
\citep{aanetal1993,frietal2004,jura2006,kilred2007,faretal2010} as has stellar dredge-up and radiative levitation, based on the effective temperature range of the surveyed WDs.

In over 35 cases, polluted WDs also harbour an observable debris disc \citep{zucbec1987,ganetal2006,faretal2009,dufetal2012, faretal2012,meletal2012,beretal2014,rocetal2014,wiletal2014,manetal2015}.
No known debris disc surrounds an unpolluted WD (see \citealt*{xuetal2015} for one potential exception), strongly suggesting that the known discs are accreting onto WDs.  
The debris discs themselves probably arise from the tidal destruction of planetesimal-like bodies
\citep{graetal1990,jura2003,debetal2012,beasok2013,veretal2014c,veretal2015b}.
Although the radial extent of the discs are well-constrained to lie within the WD's Roche (or disruption) radius, the disc mass is unconstrained.
Our best constraints on the remnant planetary mass that is disrupted and accreted instead comes from DBZ (metal-enriched helium-dominated) WDs.

DBZ WDs harbour deep convection zones which provide a record of all the mass accreted over the past Myr or so.  
The highest accreted mass is comparable to Pluto's mass \citep{giretal2012} within a Myr.
Researchers may obtain other mass estimates through the instantaneous accretion rates in DAZ 
(metal-enriched hydrogen-dominated) WDs, assuming that the accretion is in steady state.  
The Solar System's asteroid belt is about three orders of magnitude \emph{less} massive than would be necessary to reproduce these accretion rates \citep{debetal2012}; an exo-Kuiper belt is more likely to reproduce the observed \emph{rate} \citep{bonetal2011}, but has trouble reproducing the observed \emph{composition} \citep{ganetal2012,juryou2014,xuetal2014}.  

One tantalizing but so-far-unrealized frontier of extrasolar planetary science is the confirmation and characterization of exomoons 
\citep{kipping2011,simetal2012,awiker2013,lewis2013,benetal2014,kipping2014}.  
As the next largest objects after exoplanets in exoplanetary systems, exomoons could represent a vast
and massive population.  
In the Solar System, the total mass in moons 
(
$\sim6\times10^{23}\,kg$)
\footnote{http://www.wolframalpha.com/\\input/?i=mass+of+moons+in+the+solar+system}
is greater than the mass of the planets Mercury and Mars individually, and is more than two orders of magnitude greater than the total mass of the asteroid belt.

This potentially large exomoon mass reservoir has important implications for the fate of planetary systems.  
As exoplanet-hosting stars leave the main sequence (MS) and become giant branch (GB) stars, they shed between one half and four-fifths of their mass, expand their radii by many au, and increase their luminosity by up to four orders of magnitude.

Consequently, orbiting bodies are subjected to a plethora of strong forces with complex implications \citep{veras2016}.  
Exoplanets may be engulfed
into the star \citep{kunetal2011,musvil2012,norspi2013,adablo2013,viletal2014}, 
collide with each other \citep{debsig2002,veretal2013a,voyetal2013,musetal2014,vergae2015}, 
or escape the system entirely \citep{veretal2011,vertou2012,adaetal2013,veretal2014a}.
Exoasteroids instead may self-destruct
\citep{veretal2014b,veretal2015c}; those that survive may be dragged \citep{donetal2010} or perturbed 
\citep{bonetal2011,debetal2012,frehan2014,bonver2015}
into either the GB star or the resulting white dwarf (WD).  
Exo-Oort cloud comets may accrete onto the WD \citep{alcetal1986,veretal2014d,stoetal2015}, and 
second-generation planets may even be formed \citep{perets2011,beasok2014,schdre2014,voletal2014,beasok2015}.

The GB phase can also have lasting effects on the long-term dynamical stability of the planetary system.  
Instabilities do not necessarily manifest themselves
until many Gyr after the star has become a WD.  
The potential consequences for exomoons have heretofore been ignored, as the above references focus on exoplanets and/or exoasteroids.  
Remedying this neglect may help us better understand the observable signatures of late dynamical evolution in exosystems.

The total mass in exomoons is likely to be sufficiently large that moons liberated from their parent planets can play three crucial roles in post-MS systems: 
(1) to achieve an orbit around a WD which may be detectable by transit (as perhaps in WD 1145+017),
(2) to contribute directly to the polluted matter through collisions with the WD Roche radius, 
(3) to contribute indirectly by changing the orbital architecture through which smaller bodies (such as asteroids) get perturbed to the WD Roche radius.  
Consequently, quantifying the fraction of moons which escape the gravitational pull of their parent planets may represent a crucial consideration in polluted WD systems.

In this paper, we demonstrate that moons are easily liberated during the WD phase due to instabilities arising from planet-planet gravitational scattering. 
By utilizing the detailed close encounter output from the simulations in \cite{vergae2015}, we find that
(a) incursions frequently occur well within the Hill sphere of the parent planet, strongly disrupting any satellites, and in many cases, ejecting moons, and
(b) planets themselves can attain pericenters with respect to the WD of only a fraction of an AU, causing extreme Hill-sphere contraction, and the liberation of moons into WD-grazing orbits. 
Our results show that planet-planet scattering helps dissociate moons during the post-MS phases, just as on the MS phases \citep{gonetal2013}.  
Without this type of scattering, moons robustly remain bound to their parent planets, regardless of how tightly packed the planets are \citep{payetal2013}.  
We do not consider moon-moon scattering, which represents another potential vehicle for ejection \citep{perpay2014}.

In Section 2, we detail how stellar mass loss affects the stability of a moon.  
We then characterize important parameters during close encounters between planets which are scattering in Section 3.
In Section 4, we relate these parameters to the orbital excitation and escape of moons during planet-planet encounters.
In Section 5 we discuss close-pericenter approaches between planet and WD.
We discuss our results in Section 6, and conclude in Section 7.

\section{Effect of stellar mass loss on moons}
\label{SECN:ML}

First we determine how an exomoon responds to stellar mass loss from the star.  Consider a single moon orbiting a single planet, which together orbit a single star. Let $M_{\star}$, $M_{\rm p}$ and $M_{\rm m}$ represent the masses of the star, planet and moon.  Assume $M_{\star}$ is time-dependent and small enough (typically less than about $6-8M_{\odot}$) such that it will eventually become a WD.  Both $M_{\rm p}$ and $M_{\rm m}$ are considered to be fixed.

In order for the moon to orbit the planet and not the star, (i) the ratio $M_{\rm m}$/$M_{\rm p}$ must be sufficiently small, (ii) the moon-planet distance $r_{\rm m}$ must be small enough to be within the planet's gravitational sphere of influence and (iii) $r_{\rm m}$ must be large enough that the moon is outside of the planet's Roche (or disruption) radius.  

Condition (ii) is often satisfied (for prograde orbits) when $r_{\rm m} \lesssim 0.5 r_{\rm H}$, where $r_{\rm H}$ is the Hill radius

\[
r_{\rm H} \equiv 
a_{\rm p} (1 - e_{\rm p})
\left( \frac{M_{\rm p}}{3M_{\rm \star}} \right)^{1/3}
\]

\begin{equation}
\ \ \ \ 
= 2.43 \ {\rm au}
\left(
\frac{a_{\rm p}}{30 \ {\rm au}}
\right)
\left(
\frac
{M_{\rm p}}
{M_{\rm Jup}}
\right)^{1/3}
\left(
\frac
{M_{\star}}
{0.6 M_{\odot}}
\right)^{-1/3}
,
\label{Hilleq}
\end{equation}

\noindent{}and where $a_{\rm p}$ and $e_{\rm p}$ are the semimajor axis and eccentricity of the planet with respect to the star. Because this paper is focused on WDs, in the equations we adopt a fiducial WD mass of $0.6 M_{\odot}$, which corresponds to a progenitor MS mass of about $1.7 M_{\odot}$, assuming Solar metallicity \citep{huretal2000,2008ApJ...676..594K,2008MNRAS.387.1693C}.

Condition (iii) is $r_{\rm m} > r_{\rm R}$, where $r_{\rm R}$ is the Roche radius of the planet.

\[
r_{\rm R} \equiv C R_{\rm p} \left( \frac{\rho_{p}}{\rho_{\rm m}} \right)^{1/3}
\]

\begin{equation}
\ \ \ \ 
= 5.4 \times 10^{-4} {\rm au} \ \times C 
\left( \frac{M_{\rm p}}{M_{\rm Jup}} \right)^{1/3}
\left( \frac{\rho_{\rm m}}{1 \ {\rm g \ cm}^{-3}} \right)^{-1/3}
,
\label{Rocheeq}
\end{equation}

\noindent{}where $R_{\rm p}$ is the radius of the planet, and $\rho_{\rm p}$ and $\rho_{\rm m}$ are the densities of the planet and moon. The constant $C$ is dependent on the shape, spin and composition of the moon, as well as the criteria for disruption (cracking, deformation or dissociation).  Here, $C$ ranges from 0.85 to 1.89 \citep{beasok2013}.

Together, Eqs. (\ref{Hilleq}-\ref{Rocheeq}) illustrate that stable moons can orbit planets at a range of distances which span several orders of magnitude in au.  When $r_{\rm R} < r_{\rm m} \lesssim 0.5r_{\rm H}$ is satisfied,
the moon's orbit with respect to the planet can be considered fixed and stable.
As the star loses mass, (i) the planet's orbit will expand, (ii) the moon's orbit will remain unchanged (as that orbit is independent of $M_{\star}$), (iii) the value of $r_{\rm R}$ will remain unchanged, and (iv) the value of $r_{\rm H}$ will change.  

\,(i) Regarding the first point, as long as the mass loss is isotropic, the system is rotationally symmetric and angular momentum is conserved.  Consequently, with help from the vis-viva equation, the equations of motion in orbital elements may be derived \citep{omarov1962,hadjidemetriou1963,veretal2011}.  In the ``adiabatic'' case, where the averaged equations of motion, denoted by brackets, can be used (within a few hundred au; see \citealt*{veretal2011}), 

\begin{equation}
\left\langle \frac{da_{\rm p}}{dt} \right\rangle = 
- \frac{a_{\rm p}}{M_{\star} + M_{\rm p}} \frac{dM_{\star}}{dt} > 0
\label{adiaa}
\end{equation}

\begin{figure}
\centering
\includegraphics[trim = 0mm 0mm 0mm 0mm, clip, angle=0, width=\columnwidth]{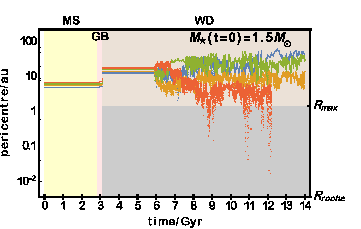}
\caption{Example of a post-MS scattering simulation, reproduced from \citet{vergae2015}, illustrating 
(a) the orbital expansion of the planetary semi-major axes (by a factor of about 2.6) due to stellar mass loss (from $1.5M_{\odot}$ to about $0.58 M_{\odot}$; see equation \ref{adiaa}) at $\sim 3$ Gyr; 
(b) the late ($>6$ Gyr) onset of planet-planet scattering discussed in Section \ref{SECN:PMS:SCAT:SAMP}, 
and 
(c) the periods of close pericenter approach discussed in Section \ref{SECN:PERI}.
Note that: 
\texttt{MS} is the Main-Sequence phase; 
\texttt{GB} is the Giant Branch phase; 
\texttt{WD} is the White Dwarf phase; 
$R_{\rm Max}$ is the star's maximum expansion radius during the GB phase, 
$R_{roche}$ is the Roche breakup radius of the WD.
}
\label{FIG:VG_F1_FIRST}
\end{figure}

\noindent{}always.  Also, on average, none of the eccentricity, inclination, argument of pericentre, or longitude of ascending node change.  Although angular momentum is no longer conserved in the anisotropic mass loss case, for realistic stars the isotropic mass loss approximation is excellent when planetary orbital separations are less than about a few hundred au \citep{veretal2013b}.
An example of planetary orbital expansion during mass loss is illustrated in Figure \ref{FIG:VG_F1_FIRST} -- reproduced from \cite{vergae2015} -- in which one can see that the planetary orbits in the pink ``GB'' strip expand by a factor of 2.6 due to the stellar mass loss from a $1.5M_{\odot}$ progenitor star.

%

\,(ii) Regarding the second point, the moon will move with respect to the planet, regardless of how the planet is changing its orbit.  
The moon will only ``feel" the mass loss from the star when the wind carrying this mass is in-between the moon and planet, effectively increasing the planet's mass\footnote{The mass accreted onto the planet itself is negligible; see Section 4b(ii) of \cite{veras2016}.}.   
In Appendix \ref{APP:WIND} we demonstrate that the maximum amount of mass within the orbit of the moon at any given time is negligible compared to the planet mass itself.

\,(iii) Regarding the third point, other forces besides mass loss (such as radiation, and erosion from the stellar wind) could in principle change the physical state of the planet and moon.  
Consequently, the coefficient $C$ in equation (\ref{Rocheeq}) might undergo a slight change.  
But largely the densities of the moon and planet remain unchanged, and hence $r_{\rm R}$ remains unchanged too.

\,(iv) Finally, for the fourth point regarding the change in Hill radius, we combine the standard equation for the Hill radius (equation \ref{Hilleq}) with the isotropic mass loss equations \citep{hadjidemetriou1963} to obtain

\[
\frac{dr_{\rm H}}{dt} = 
-r_{\rm H}(a_{\rm p})
\left(
\frac{M_{\rm p}}{3M_{\rm \star}}
\right)
\frac{dM_{\star}}{dt}
\]

\begin{equation}
\ \ \ \ \ \ \ \ \
\times
\frac
{M_{\rm p} + M_{\star} \left(4 - 3\cos{f}\right) 
+ \left(M_{\star} + M_{\rm p}\right)e_{\rm p}}
{M_{\rm p} \left(M_{\rm p} + M_{\star}\right)\left(1+e_{\rm p}\right) }
.
\label{EQN:drH_dMgen}
\end{equation}

\noindent{}Because the Hill radius is derived assuming that
$M_{\rm p} \ll M_{\star}$, we can write

\begin{equation}
\frac{dr_{\rm H}}{dt} 
\approx -r_{\rm H}(a_{\rm p})
\left[
\frac{4 - 3 \cos{f} + e_{\rm p}}{3\left(1+e_{\rm p}\right)}
\right]
\frac{1}{M_{\star}}
\frac{dM_{\star}}{dt}
.
\label{EQN:drH_dMsimp}
\end{equation}

\noindent{}Both Equations \ref{EQN:drH_dMgen} and \ref{EQN:drH_dMsimp} demonstrate that the \emph{direct} effects of central stellar mass-loss cause $dr_{\rm H}/dt > 0$ always (because $dM_{\star}/dt<0$)\footnote{But in Section \ref{SECN:PERI} we examine indirect effects which can cause temporarily cause $dr_{\rm H}/dt < 0$ during close pericenter passages between planet and WD.}.  
Consequently, because a moon's orbit with respect to its parent planet remains fixed, {\it moons become more stable due to stellar mass loss alone}.  
After post-MS mass loss, the value of $r_{\rm m}/r_{\rm H}$ has been lessened.

We can estimate the change in $r_{\rm m}/r_{\rm H}$ by considering the averaged (adiabatic) mass loss equations of motion.  Assume that the final WD mass is $M_{\star}^{\rm WD}$ and the initial progenitor stellar mass is $M_{\star}^{\rm MS}$. In this case,

\begin{equation}
r_{\rm H}^{\rm WD} 
\approx
a_{\rm p}^{\rm MS} \left(1 - e_{\rm p}^{\rm MS}\right)
\left( \frac{M_{\rm p}}{3M_{\star}^{\rm WD}} \right)^{1/3}
\left( \frac{M_{\star}^{\rm MS}}{M_{\star}^{\rm WD}}\right)
\end{equation}

\noindent{}Consequently,

\begin{equation}
\frac{r_{\rm H}^{\rm WD} }{r_{\rm H}^{\rm MS} }
=
\left(
\frac{M_{\star}^{\rm MS} }{M_{\star}^{\rm WD} }
\right)^{4/3}
\label{eqnrh}
\end{equation}

\noindent{}We obtain intuition for the value of $M_{\star}^{\rm WD}/M_{\star}^{\rm MS}$ by creating stellar tracks with the {\sc SSE} code \citep{huretal2000}.  Assuming Solar metallicity, a Reimers mass loss coefficient of 0.5 on the red GB phase, and the code's default superwind prescription on the asymptotic GB phase, we find for $M_{\star}^{\rm MS} = \left\lbrace 8,7,6,5,4,3,2,1 \right\rbrace M_{\odot}$ that
$r_{\rm H}^{\rm WD}/r_{\rm H}^{\rm MS} = \left\lbrace
9.83, 9.53, 9.16, 8.55, 7.65, 6.35, 4.57, 2.39
\right\rbrace$.  Hence, the Hill radius increases by a factor of 2-10 due to post-MS mass loss, entrenching the moons deeper within the Hill radius.  
Consequently, liberating moons from this more secure position requires violent close encounters.


\section{Scattering velocities and impact parameters}
\label{SECN:SCATT_VEL}

Having demonstrated that stellar mass loss entrenches moons deeper within the Hill radius of the parent planet, we now consider the susceptibility of these moons to close encounters with other planets.  Although the results in this section are specific to WD systems, they may provide insight into more general planet-planet scattering studies.

\emph{ In this section we examine the spectrum of \underline{multiple} close-encounters experienced by planets over the course of billions of years of evolution, drawing on simulations of a variety of different system masses and architectures.}

\subsection{Post-MS scattering sample}
\label{SECN:PMS:SCAT:SAMP}

We obtain typical scattering velocities and impact parameters by using the data from the simulations performed in \cite{vergae2015}, who evolved packed systems of 4 and 10 planets throughout all phases of stellar evolution post-formation and demonstrated that instability can {\it first} occur during the WD phase.  That paper extended previous studies modeling the post-MS evolution of two-planet \citep{veretal2013a} and three-planet \citep{musetal2014} systems, but with a set of progenitor masses ($1.5M_{\odot}$-$2.5M_{\odot}$) which better reflect the currently-observed WD population \citep[Fig. 1 of][]{koeetal2014}.

\cite{vergae2015} simulated five basic types of planetary system:
(i) Systems of four Jupiter-mass planets, with the innermost planet initially at $5\,au$;  
(ii) \& (iii) Systems of four Earth-mass planets, with the innermost planet initially at $2\,au$ and $5\,au$ respectively; 
and
(iv) \& (v) Systems of ten Earth-mass planets, with the innermost planet initially at $5\,au$ and $10\,au$ respectively.

We consider only those simulations from \cite{vergae2015} in which planetary systems first unpacked (became unstable) after the end of the MS. 
The fraction of systems becoming unstable varies according to system architecture: we refer the reader to \cite{vergae2015} for further details.
An example of such a simulation is reproduced in Figure \ref{FIG:VG_F1_FIRST}, illustrating the late onset of instability (about 3 Gyr after the star has become a WD).
While planet-planet scattering can occasionally occur on the main-sequence (e.g. \citet{1996Sci...274..954R}), potentially removing exomoons, the results of \cite{vergae2015} demonstrate that scattering can \emph{begin} during the post-MS phase, ensuring that for such late-scattering systems, no exomoons will have been removed throughout the MS and GB phases.

For these simulations, all close encounters between planets within $3r_{\rm H}$ were recorded.  
The information obtained in these recordings were the pericentre distance, $q$, and the velocities of the planets at their closest approach.  
We denote the relative speed at this closest approach as $V_{q}$.

During a close encounter, the planets are on hyperbolic orbits with respect to each other.  Consequently, for this orbit we can define an impact parameter $b$ and an ``initial" velocity $V_{\infty}$ through \citep[see eqs. 3 and 6 of][]{vermoe2012}

\begin{equation}
b = q \left[1 - \frac{2\mu}{q V_{q}^2}  \right]^{-1/2}
\approx
q + \frac{\mu}{V_{q}^2}
\approx
q
,
\label{beqq}
\end{equation}

\begin{equation}
V_{\infty} = \sqrt{V_{q}^2 - 2 \frac{\mu}{q}}
\approx
V_{q}
,
\label{infeqq}
\end{equation}

\noindent{}where $\mu \equiv G\left(M_{\rm p} + M_{\rm f}\right)$, such that $M_{\rm f}$ is the mass of the ``flyby" planet (the planet not hosting a moon), and the approximations employed hold for the typical range of planet-planet encounter parameters illustrated in Figures \ref{FIG:OCC:B} and \ref{FIG:SCATT_DIST_A}.

\subsection{Distribution of Scattering Parameters}
\label{SECN:PMS:SCAT:PARAM}
%
%

\begin{figure*}
\centering
\parbox{\textwidth}{
\includegraphics[trim = 2mm 2mm 2mm 2.4mm, clip, angle=0, width=0.95\textwidth]{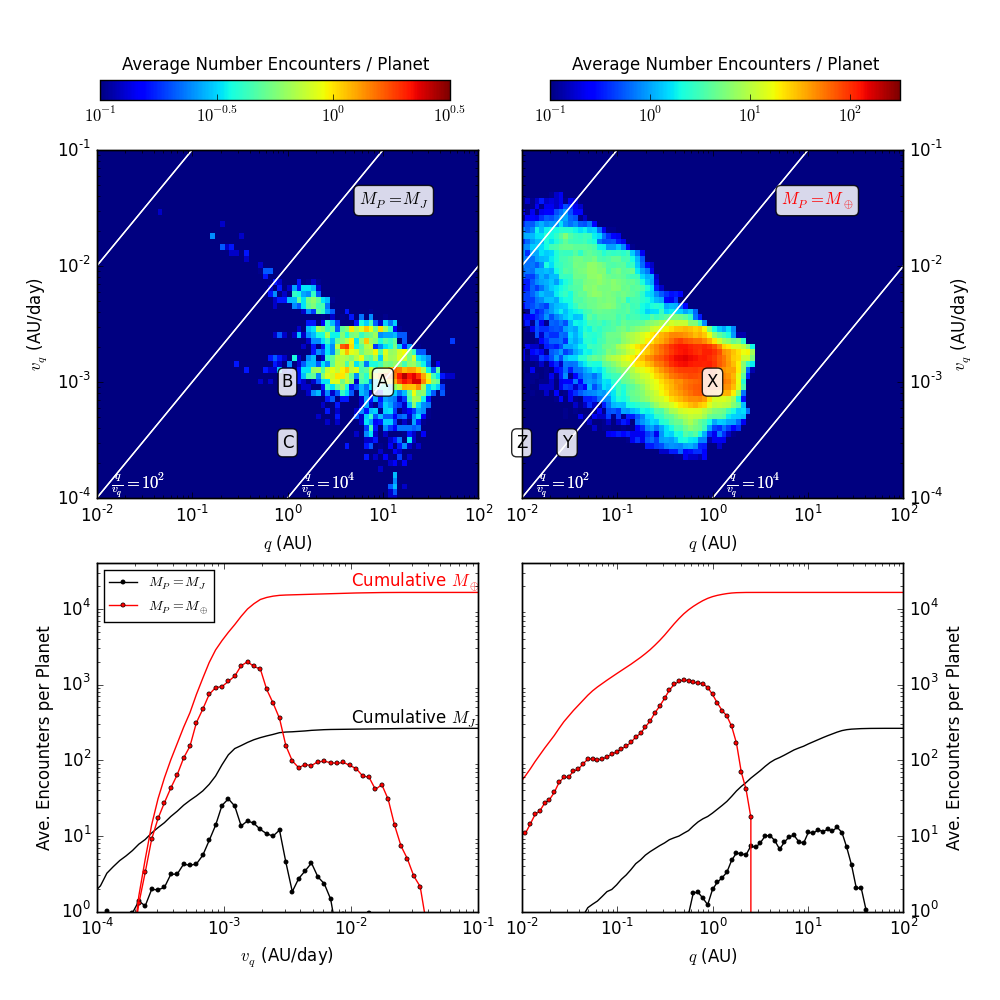}
\caption{
Distribution of close-approach parameters for Jupiter-mass Planets (top-left), and Earth-mass planets (top-right).
These systems have $N_P=4$ equal-mass planets initialized such that the innermost planet (prior to mass-loss) is at $5$ au.
The histograms for the same systems are plotted below, illustrating the \emph{average number of encounters per planet} with $v_q$ (bottom-left) and $q$ (bottom-right). 
Note that both histograms and cumulative curves are plotted.
The systems of Earth-mass planets experience far greater numbers of encounters, primarily because these systems do \emph{not} eject planets (whereas the Jupiter-mass systems do), allowing the Earth-mass planets to experience multiple close-approaches over the age of the system.
In Figure \ref{FIG:SCATT_DIST_A} below, we provide additional cumulative histogram curves. 
The labels $A$,$B$,$C$,$X$,$Y$, and $Z$ correspond to the detailed simulations illustrated in Section \ref{SECN:MOONS} and Figure \ref{FIG:MOON:SURVIVE}, where we examine the loss of moons during individual close planet-planet encounters 
}
\label{FIG:OCC:B}
}
\end{figure*}
%
%
%
%
\begin{figure*}
\begin{minipage}[b]{\textwidth}
\centering
\begin{tabular}{c}
\includegraphics[trim = 0mm 0mm 0mm 0mm, clip, angle=0, width=0.9\textwidth]{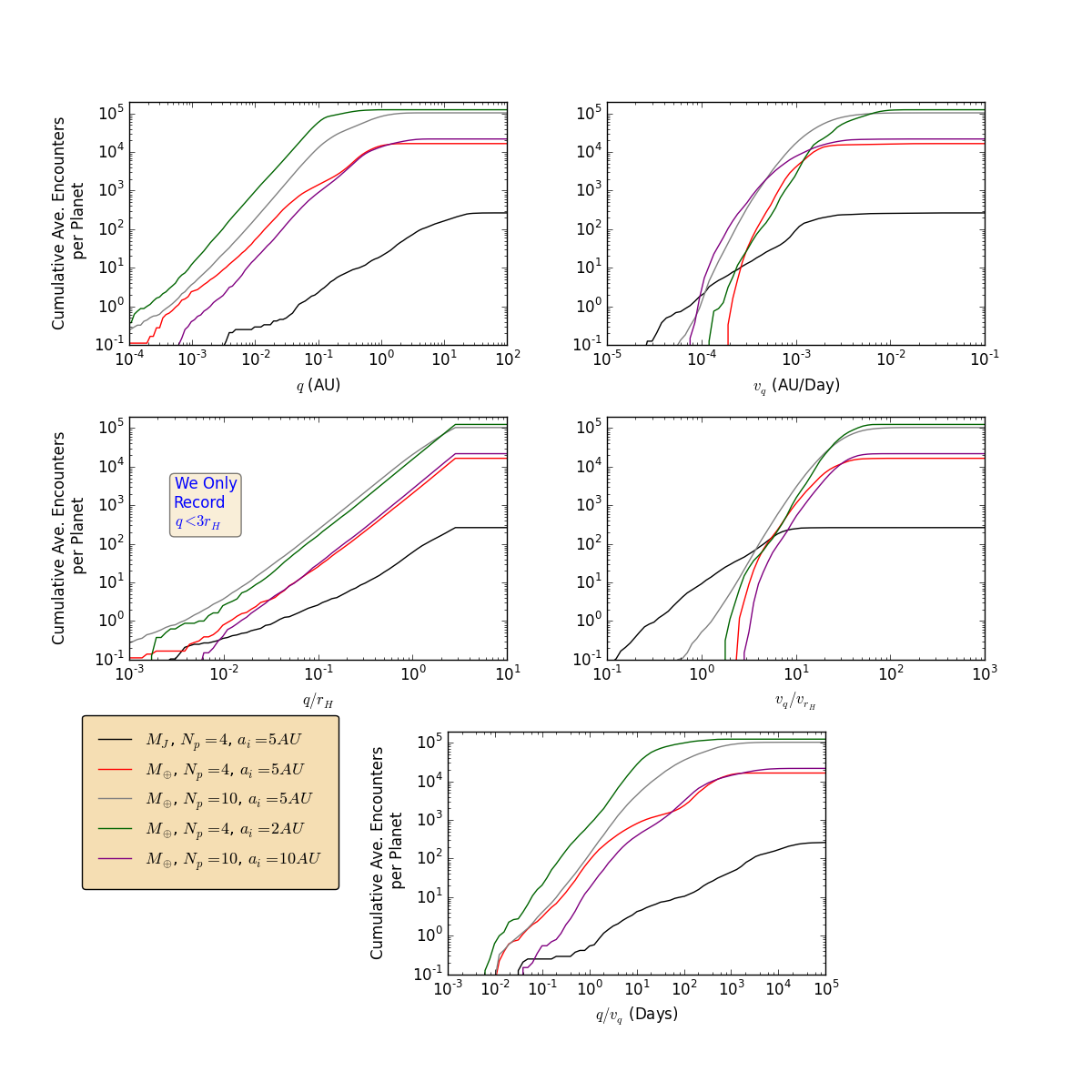}
\end{tabular}
\caption{Distribution of close-approach parameters for different sets of planet-planet scattering simulations.
Cumulative histograms for:
{\bf Top-Left} $q$;
{\bf Top-Right} $v_q$;
{\bf Mid-Left} $q/r_{\rm H}$;
{\bf Mid-Right} $v_q/v_{R_H}$;
{\bf Bottom} $q/v_q$;
The colors denote the different initial conditions:
\emph{Black:}  $M_J,\,       N_P=4, \,a_i=5$ au; 
\emph{Red:}    $M_{\oplus},\,N_P=4, \,a_i=5$ au; 
\emph{Gray:}   $M_{\oplus},\,N_P=10,\,a_i=5$ au; 
\emph{Green:}  $M_{\oplus},\,N_P=4, \,a_i=2$ au; 
\emph{Purple:} $M_{\oplus},\,N_P=10,\,a_i=10$ au; 
We note that the Black and Red curves above for $q$ and $v_q$ are repeated from Figure \ref{FIG:OCC:B} above. 
We see that on average, 
(i) a Jupiter-mass planet around a post-MS WD which undergoes late instability will experience $\sim 100$ close encounters with $q\,<\,3\,r_{\rm H}$, of which $\sim 10$ have a timescale $\left(q / v_q\right) \sim\,10$ days, while
(ii) a system of unstable Earth-mass planets will experience $> 10^4$ close encounters with $q\,<\,3\,r_{\rm H}$, of which $10\,-\,10^4$ have a timescale $\left(q / v_q\right) \sim\,1$ day (i.e. there is a broad range, dependent on initial conditions).
}
\label{FIG:SCATT_DIST_A}
\end{minipage}
\end{figure*}

We plot the cumulative distributions of $q$ and $V_{q}$ across all simulations in Figures \ref{FIG:OCC:B} and \ref{FIG:SCATT_DIST_A}.  
These curves provide insight into the dynamics of close encounters, and are typically not featured in dedicated exoplanet scattering studies. 
The distributions of $b$ and $V_{\infty}$ are visually almost indistinguishable from those of $q$ and $V_{q}$ and hence are not shown.

These figures demonstrate that the encounters are penetrative; the minimum value of $q$/$r_{\rm H}$ is such that on average, in systems of Earth-mass planets, each planet experiences at least one close-approach with $q$/$r_{\rm H} \lesssim 10^{-2}$, while in systems of Jupiter-mass planets, each planet experiences at least one close-approach with $q$/$r_{\rm H} \lesssim 5\times 10^{-2}$.

We note that the Earth-mass planets experience nearly two orders-of-magnitude more close-encounters than do the Jupiter-mass planets. 
This large difference essentially arises because the Earth-mass planets don't have the energy to eject one another, so they are doomed to experience repeated close-encounters unless they eventually collide, where-as the Jupiter-mass planets can be entirely ejected from the system, curtailing the number of close planetary encounters.

\section{Effect on moons: Numerical Scattering Experiments}
\label{SECN:MOONS}
Having illustrated the distribution of parameters during the close approaches amongst planets in post-MS exosystems, we now consider how destructive these encounters are to orbiting moons.  

We perform numerical simulations which model the evolution of moons after the close encounters experienced by the systems simulated in \cite{vergae2015}.
In Appendix \ref{SECN:MOONS:ANALYTIC} we take an analytical approach and determine in what regimes might an impulse approximation be applicable and able to explain our numerical results.

\emph{ In this section we examine the effect of a \underline{single} close-encounter from the numerous such encounters demonstrated to occur in Section \ref{SECN:SCATT_VEL}. }

\subsection{Methodology for Integration with {\sc Mercury}}
\label{SECN:MOONS:NUMERICAL:METHOD}
Our basic physical scenario consists of a parent planet in an orbit with $a_{\rm p}=30$ au and $e_{\rm p}=0$ around a central star of mass $1M_{\odot}$.
We set the mass of the parent planet at either $M_{\rm p}=M_J$ or $M_{\rm p}=M_{\oplus}$ (see Table \ref{TAB:MOONS} for simulation parameters).

\begin{table*}
\begin{minipage}[b]{\textwidth}
\centering
\caption{Parameter variations in numerical integrations of moon perturbations due to fly-by encounters with another planet. We set
$M_{\star}=1M_{\odot}$, $a_{\rm p}=30$ au, and $e_{\rm p}=0$. 
The period of the planet is $\sim 164\,$yrs, hence the period of a moon at $a_{\rm m}=0.5r_{\rm H,p}$ is $\sim 33.5\,$ yrs or $1.2\times10^4\,$days.
The total number of simulated moons, $n_{\rm m}=10^4$.
The initial planet-planet separation equals $4r_{\rm H}$.
The close-encounter pericenter is $q$ (au) and the velocity at pericenter is $V_{\rm q}$ (au/day).
The total simulation time is $t_{\rm sim}$ (days).
Additional definitions can be found in Table \ref{TAB:DEFN}}
%
%
\label{TAB:MOONS}
\begin{tabular}{l ccccc l } 
\hline
Simulation 	 &	 $q$ 	        &	 $V_{\rm q}$ 	            &	 $M_{\rm p}$    &	$r_{\rm H}$ 	 &	 $t_{\rm sim} $             &   \\
Set 	     &	 $({\rm au})$   &	 $({\rm au}/{\rm day})$     &	  	            &	$({\rm au})$ 	 &	 $({\rm days}) $            &   Notes   \\
\hline
\hline
$\texttt{A}$     &       $1.0 \times 10^{1}$     &       $1.0 \times 10^{-3}$    &       $M_J$   &       2.0     &       $2.5 \times 10^{4}$         & Common, Little effect\\
$\texttt{B}$     &       $1.0           $        &       $1.0 \times 10^{-3}$    &       $M_J$   &       2.0     &       $2.5 \times 10^{4}$         & Relatively Rare, Destructive\\
$\texttt{C}$     &       $1.0           $        &       $3.0 \times 10^{-4}$    &       $M_J$   &       2.0     &       $8.3 \times 10^{3}$         & Very Rare, Highly destructive\\
$\texttt{X}$     &       $1.0           $        &       $1.0 \times 10^{-3}$    &       $M_{\oplus}$    &       0.3     &       $3.6 \times 10^{3}$ & Common, Little effect\\
$\texttt{Y}$     &       $3.0 \times 10^{-2}$    &       $3.0 \times 10^{-4}$    &       $M_{\oplus}$    &       0.3     &       $1.2 \times 10^{4}$ & Relatively Rare, Destructive\\
$\texttt{Z}$     &       $1.0 \times 10^{-2}$    &       $3.0 \times 10^{-4}$    &       $M_{\oplus}$    &       0.3     &       $1.2 \times 10^{4}$ & Very Rare, Highly destructive\\
\hline
\end{tabular}
\end{minipage}
\end{table*}

The parent planet was initialized with a swarm of $n_{\rm m}=10^4$ test-particle moons ($M_{\rm m}=0$).
The moons are initialized with:
(i) semi-major axes drawn from a uniform distribution in log-space with $R_{\rm p} < a_{\rm m} / r_{\rm H} < 0.5 $, and (ii) eccentricities drawn from a uniform linear distribution between 0 and 1.
The semi-major axes and eccentricity were then jointly constrained to ensure that the initial conditions have pericenter $ > R_{\rm p}$ and apocenter $ < 0.5r_{\rm H,p}$.
An example of these initial conditions in $(a_{\rm m},e_{\rm m})$-space is plotted in Figure \ref{FIG:MOON:SURVIVE}.
The inclinations were drawn randomly from a uniform distribution $0<i_{\rm m}<180^{\circ}$, while $\Omega_{\rm m}$, $\omega_{\rm m}$ and $\texttt{M}_{\rm m}$ were all drawn from a uniform distribution between $0^{\circ} - 360^{\circ}$.

A second planet, the ``fly-by'' planet, is injected into the simulation on a trajectory that results in a close-encounter with the parent planet, with a pericenter of $q$ and a velocity at pericenter of $V_{\rm q}$.
The fly-by planet has mass $M_{\rm f}=M_{\rm p}$.

We set the initial pairwise planet separation distance at $4r_{\rm H}$, then integrated the planets through the close encounter (with the given values of $q$ and $V_{\rm q}$) and continued the integration until the fly-by planet receded to a distance $\approx 4r_{\rm H}$ from the parent planet.

On a practical note, finding the initial relative positions and velocities required to achieve a given $q$ and $v_q$ at pericenter is slightly non-trivial. 
We could have tried to use an approach similar to that in \citet{vermoe2012}, but found the heliocentric orbital arcs to be problematic.
Hence we took the simple approach of (i) setting the planets up at pericenter (close-approach) with the desired $q$ and $V_{\rm q}$ (with no moons at this point), (ii) integrating them backwards (by reversing the relative velocity vector of the fly-by planet to that of the parent planet) until their separation was $\sim4r_{\rm H}$, (iii) added the moons to the parent planet, and finally (iv)  re-reversed the velocity of the fly-by planet at that point to run them forwards in time through the close-encounter. 

The integrations were performed using the Bulirsch-Stoer algorithm from the {\sc Mercury} $N$-Body package of \citet{Chambers99}.

\subsection{Parameters Explored during Integrations}
\label{SECN:MOONS:NUMERICAL:VARIATIONS}
We initialized our simulations such that they resulted in close-encounters with a range of $q$ and $V_{\rm q}$, and used the distribution of close-approach parameters in Figure \ref{FIG:OCC:B} as a guide to the appropriate range to cover. 
In Table \ref{TAB:MOONS} we provide a detailed list of the close-approach simulations performed and the key parameter variations for each.

\subsection{Results of Numerical Integrations}
\label{SECN:MOONS:NUMERICAL:RESULTS}
\begin{figure*}
\begin{minipage}[b]{\textwidth}
\centering
\begin{tabular}{cc}
\includegraphics[trim = 0mm 0mm 20mm 5mm, clip, angle=0, width=0.4\textwidth]{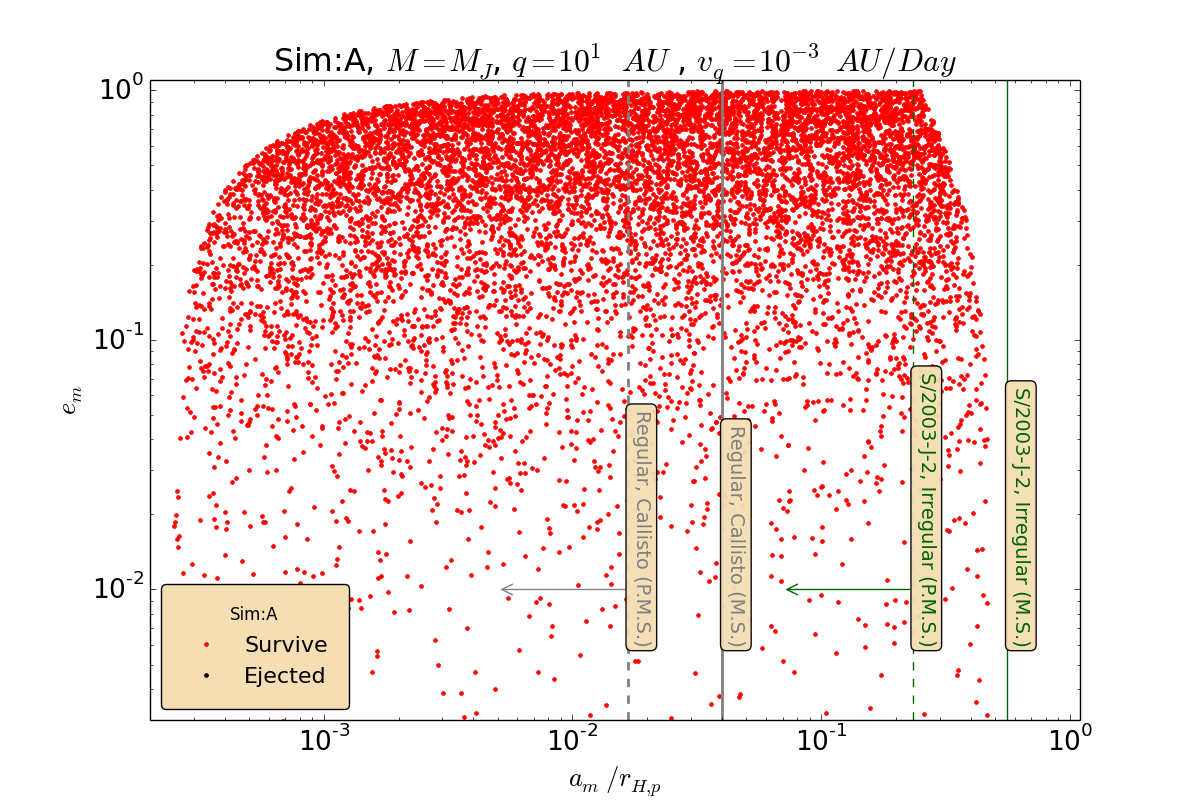}&
\includegraphics[trim = 0mm 0mm 20mm 5mm, clip, angle=0, width=0.4\textwidth]{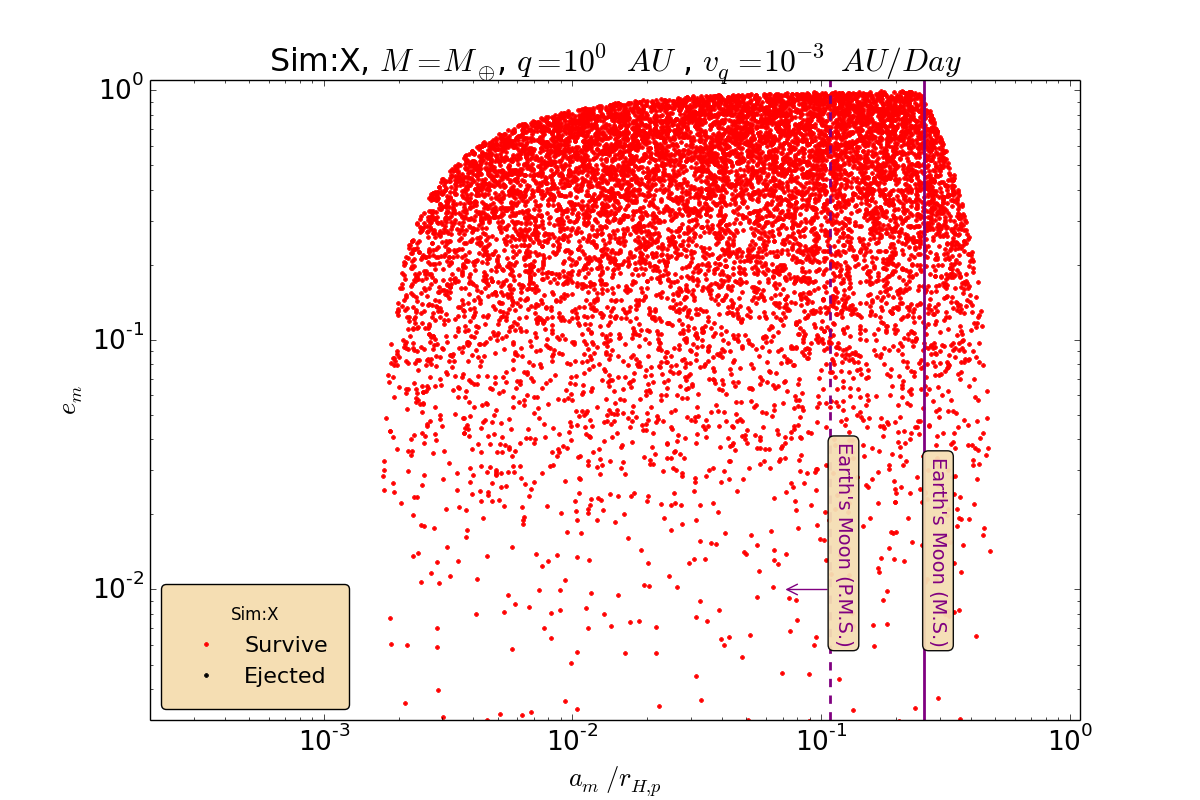}\\
\includegraphics[trim = 0mm 0mm 20mm 5mm, clip, angle=0, width=0.4\textwidth]{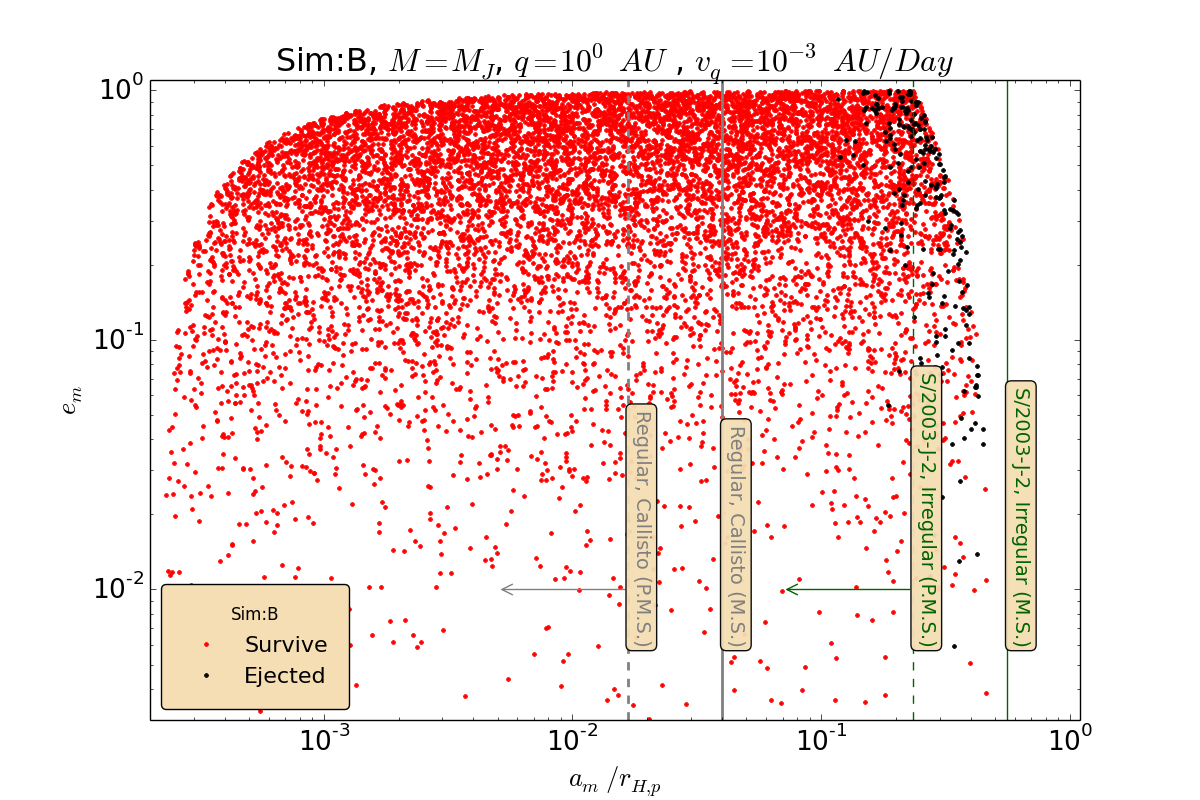}&
\includegraphics[trim = 0mm 0mm 20mm 5mm, clip, angle=0, width=0.4\textwidth]{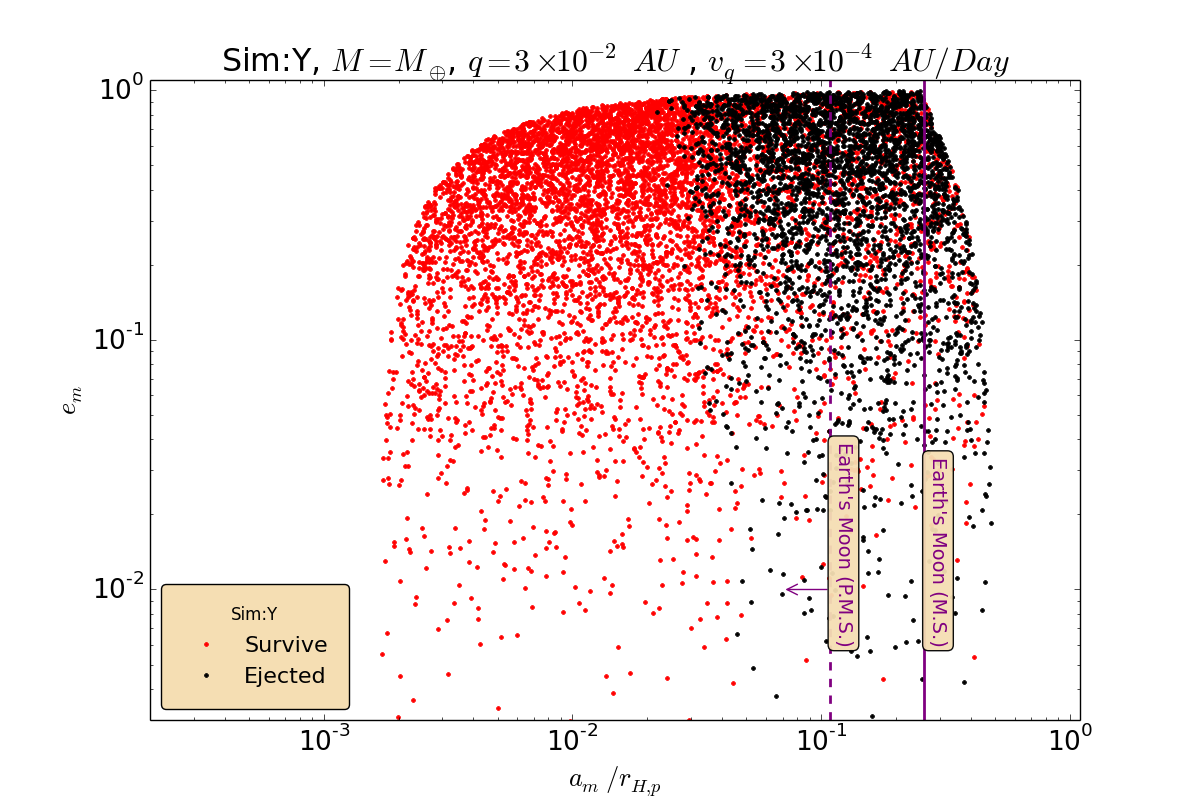}\\
\includegraphics[trim = 0mm 0mm 20mm 5mm, clip, angle=0, width=0.4\textwidth]{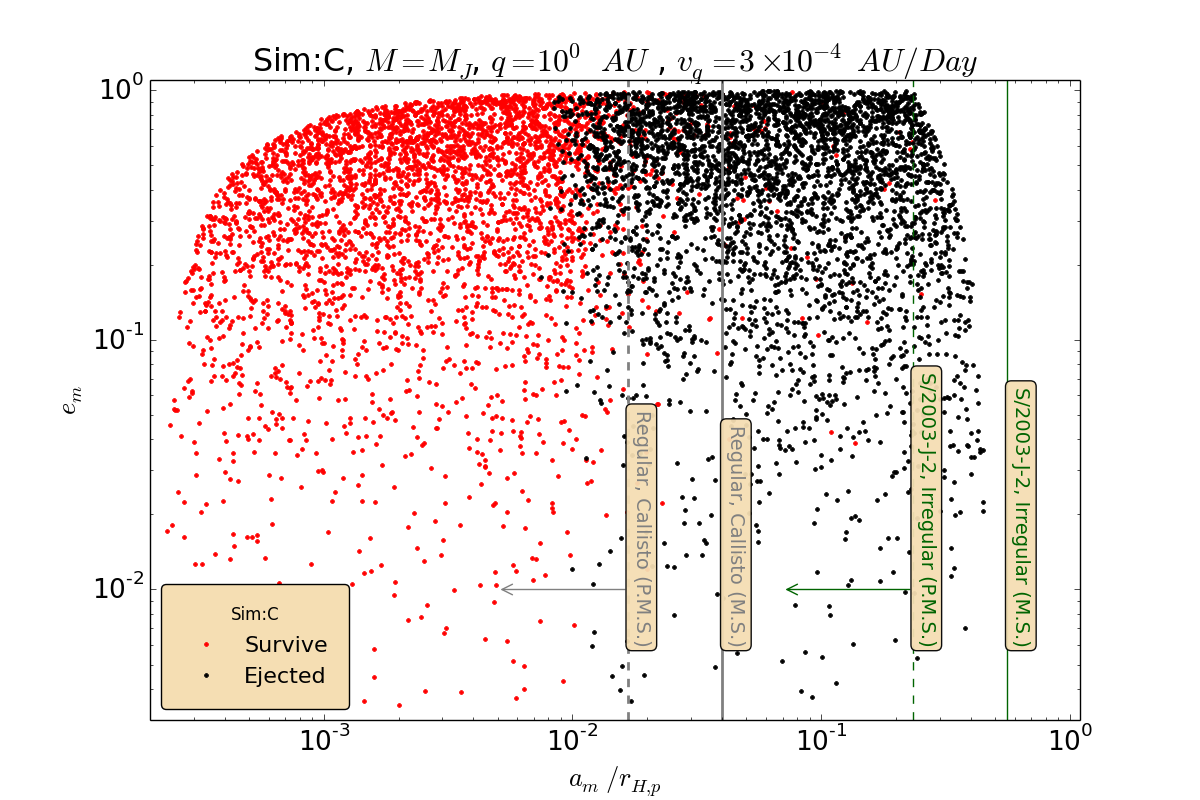}&
\includegraphics[trim = 0mm 0mm 20mm 5mm, clip, angle=0, width=0.4\textwidth]{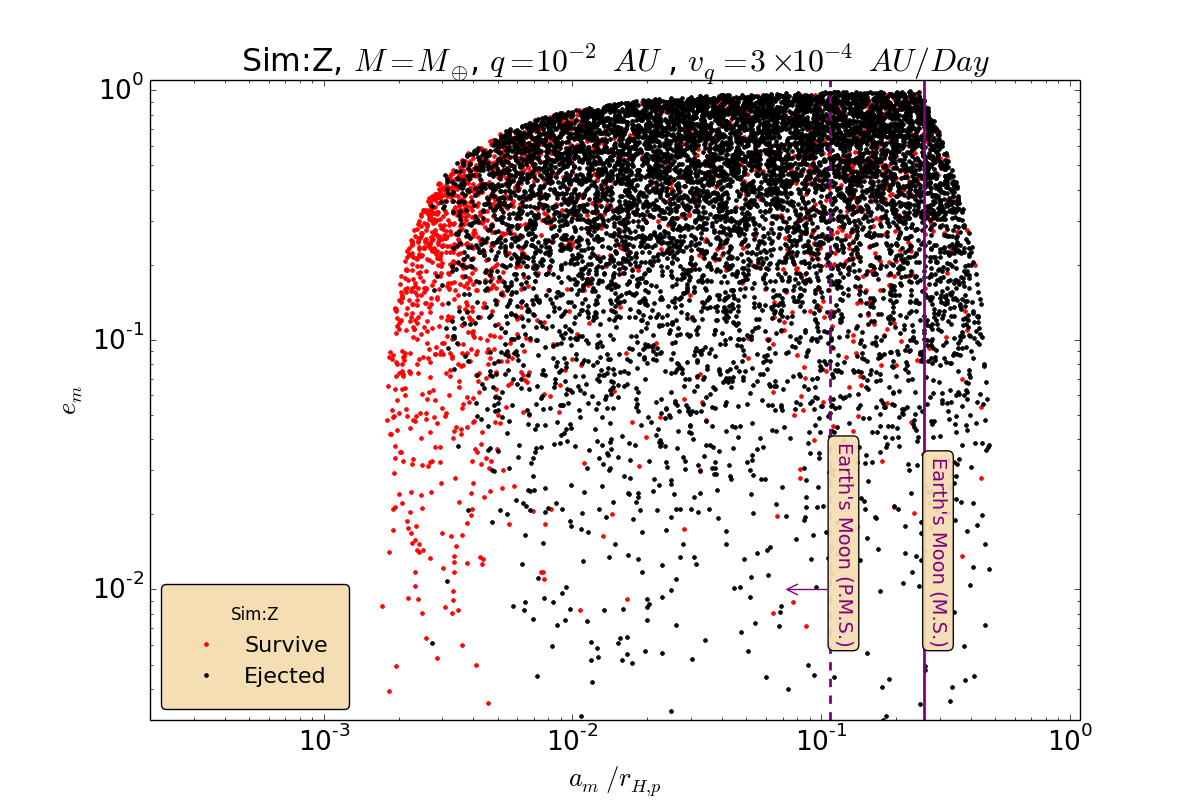}\\
\end{tabular}\\
\includegraphics[trim = 0mm 5mm 0mm 10mm, clip, angle=0, width=0.65\textwidth]{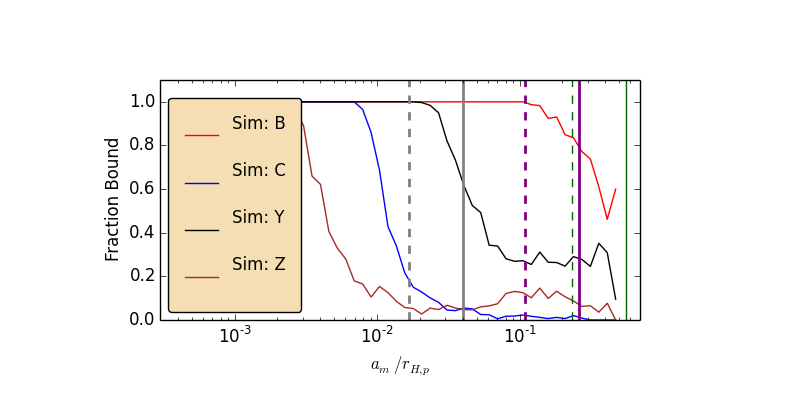}
\caption{%
Plots of the initial values of $(a_{m}/r_{H,p},e_{m})$ for $10^4$ moons in each the close-encounter simulations from Table \ref{TAB:MOONS}. 
The red dots denote moons which survive the encounter and remain bound.
The black dots are moons which were ejected.
The panels on the left are Jupiter-mass planets
The panels on the rights are Earth-mass planets. 
To guide the eye, we use solid vertical lines to plot the current semi-major axes of the outer regular moon Callisto (gray), the outer irregular moon S/2003-J-2 (green) and The Moon (purple).
We then use dashed lines to plot their relative semi-major axis after the post-MS stellar mass loss (assuming no tidal evolution), and add arrows to indicate that these limits can be further inward, depending on the degree of stellar mass loss (see Section 2).
It is clear that common but \emph{distant} encounters of the type simulated in $A$ (top-left) and $X$ (middle-left) have little effect on the moons, while the closer, but less common (each planet should experience $\sim 1$) encounters simulated in $B$ (top-right) and $Y$ (middle-right) are disruptive, while very close and very rare encounters such as $C$ and $Z$ are very disruptive.
The bottom panel illustrates the survival fraction as a function of semi-major axis, demonstrating that the \emph{single} close encounter illustrated in $B$ and $Y$ cause $\gsim50\%$ of satellites to be ejected over a large volume of parameter space.
}
\label{FIG:MOON:SURVIVE}
\end{minipage}
\end{figure*}

In Figure \ref{FIG:MOON:SURVIVE} we plot the initial values of $(a_m/r_{H},e_m)$ of $10^4$ moons in each the six close-encounter simulations described in Table \ref{TAB:MOONS}. 
The variable $r_{H}$ represents the Hill Radius of the parent planet. 
We plot in red those orbits which survive the encounter and remain bound, and in black those moons which were ejected.
The panels on the left are for Jupiter-mass planets, while those on the right  are for Earth-mass planets. 

It is clear that common but \emph{distant} encounters of the type simulated in $A$ (top-left) and $X$ (top-right) have little effect on the moons, while the closer, but less common encounters simulated in $B$ (middle-left) and $Y$ (middle-right) are more disruptive. 
Very rare encounters of the type plotted in $C$ and $Z$ (bottom-left and bottom-right respectively) can eject moons from the vast majority of the Hill sphere from a single close encounter. 
We emphasize that although the encounters in $B$ and $Y$ are indeed less-common, they were deliberately selected so that each planet experiences on average at least $1$ such encounter that was at least this destructive (see Figure \ref{FIG:OCC:B}). 

From the histogram at the bottom of Figure \ref{FIG:MOON:SURVIVE}, one can see that \emph{single} close encounters of the type in Simulations $B$ and $Y$ cause $\gsim50\%$ of satellites to be ejected from a large volume of the Hill sphere.
Such encounters would clearly be sufficient to eject the majority of loosely bound irregular moons, as well as much more massive objects such as our own moon.

We emphasize that many of the results obtained herein will apply directly to MS stars as well as the post-MS WDs we have focused on.
In other words, planet-planet scattering around MS stars will cause the loss of moons \citep{gonetal2013}.
A more detailed investigation of the parameter dependence is warranted in order to gain an understanding of 
(a) the cumulative effect of multiple close-encounters (Figure \ref{FIG:MOON:SURVIVE} illustrates the effect of only a single encounter out of the many seen in Figure \ref{FIG:SCATT_DIST_A})
, and, 
(b) when/if moons can survive the planet-planet scattering process at all stages of the stellar life-cycle. 

In Appendix \ref{SECN:MOONS:ANALYTIC} we provide some analytic approximations to determine in what regimes might an impulse approximation be applicable and able to explain our numerical results.

\section{Distribution of Planetary Pericenters (with respect to the White Dwarf)}
\label{SECN:PERI}
In Figure \ref{FIG:VG_F1_FIRST} we find that during the scattering process, the innermost planet occasionally attains a pericenter (with respect to the central WD) as low as $\sim 0.01$ au.
This has important ramifications for any moons still orbiting the planet at the point of close-pericenter approach. 
Even if no moons have been lost during close planet-planet encounters of the type modelled in Section \ref{SECN:MOONS}, at the time of pericenter passage, the radius of the planetary Hill sphere will shrink significantly, as $r_{\rm H}\propto\,q$ (see Eqn \ref{Hilleq}).
This shrinkage will cause any outer moons with $a_{\rm m} > r_{\rm H}(q)$ to be lost as the Hill radius contracts.

\begin{figure*}
\begin{minipage}[b]{\textwidth}
\centering
\begin{tabular}{c}
\includegraphics[trim = 0mm 0mm 20mm 5mm, clip, angle=0, width=\textwidth]{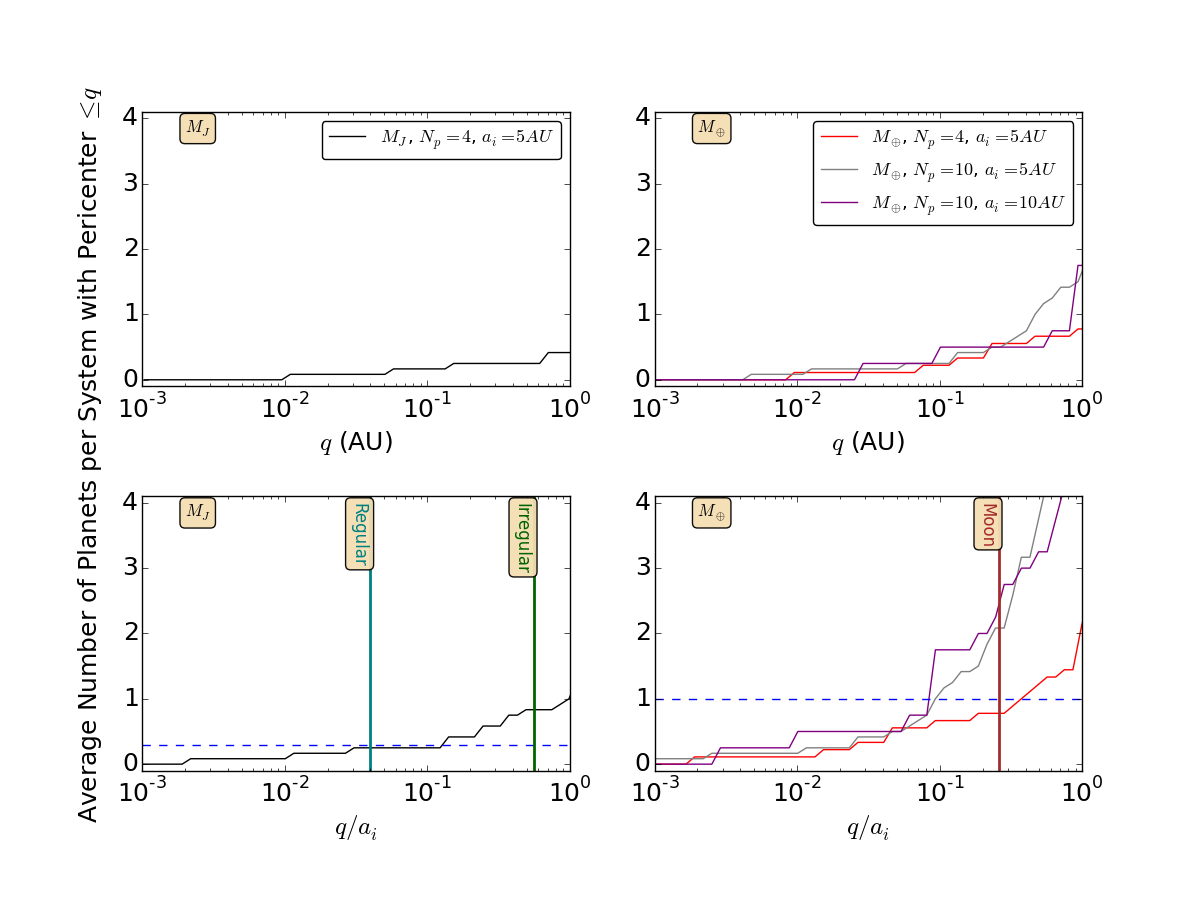}
\end{tabular}
\caption{%
Average number of planets per system with a given close-approach pericenter.
{\bf Left:} Jupiter-Mass planets.
{\bf Right:} Earth-Mass Planets.
{\bf Top:} Results as a function of absolute pericenter (au).
{\bf Bottom:} Results as a function of pericenter scaled by the initial semi-major axis of the inner planet.
The dotted lines illustrated values of 0.33 (Jupiter-mass planets) and 1.0 (Earth-mass planets) respectively, and are merely to guide the eye. 
We find that around a third of Jupiter-mass systems will experience a planet scattering to $q/a_i \leq 0.1$, while every Earth-mass simulation on average experiences at least one planet scattering to $q/a_i \leq 0.1$.
Hence, the majority of irregular moons would be lost from Jupiter-mass systems, while The Moon would be lost from Earth-mass systems.
}
\label{FIG:PERI}
\end{minipage}
\end{figure*}

To understand the frequency with which close-pericenter approaches occur, we plot in Figure \ref{FIG:PERI} the distribution of close pericenter approaches seen in the scattering simulations of Section \ref{SECN:PMS:SCAT:SAMP}.
We present the results as functions of both the absolute pericenter, $q$, and as functions of the pericenter scaled by the initial semi-major axis of the inner planet, $q/a_i$. 
The latter is used to indicate the degree by which the Hill radius of the planet will have shrunk compared to the initial scale of the Hill radius before any mass-loss, or planet-planet scattering took place. 

Given the unknown distribution of orbital parameters for exomoons, we consider a nominal value of $q/a_i \approx 0.1$, indicating that the Hill radius has shrunk by a factor of 10, or that the Hill volume has shrunk by a factor of 1,000. 
Such a reduction would suffice to unbind the Earth's Moon, many of the irregular moons of Jupiter and Saturn, and would be within a factor $\sim 2$ of unbinding outer regular moons such as Callisto.
I.e. it is a significant reduction that would act to liberate many moons in the Solar System. 

We find that around a third of Jupiter-mass systems will experience a planet scattering to $q/a_i \leq 0.1$, while every Earth-mass simulation on average experiences at least one planet scattering to $q/a_i \leq 0.1$.
These frequencies would cause the majority of irregular moons to be lost from the close-approach planet in Jupiter-mass systems, while the loss of \emph{regular} moons at smaller $a_{\rm m}$ would be less common. 
Around systems of Earth-mass planets, The Moon would be lost at such pericenter approaches. 

It is unclear from Figures \ref{FIG:MOON:SURVIVE} and \ref{FIG:PERI} whether the close planet-planet scatterings seen in Figure \ref{FIG:MOON:SURVIVE} are more efficient at liberating exomoons than the close-pericenter passages illustrated in Figure \ref{FIG:PERI}.
Of particular importance will be understanding the number (and properties) of close planet-planet scatterings which occur \emph{before} the first close-pericenter passage with the WD, as this will dictate whether any exomoons remain bound to the planet, ready to be ejected into a WD-grazing orbit. 

However, we note that exomoons liberated during planet-planet scatterings will typically be at \emph{large} distances from the WD, and their subsequent fate will presumably entail repeated chaotic scatterings (as they are on planet-crossing orbits), making it unclear just what fraction might ultimately be delivered to the central WD. 
In contrast, the moons liberated at close pericenter-approach to the WD \emph{may} be fewer in number, but importantly, when moons are lost from the planet, they will automatically occupy orbits about the WD \emph{with a pericenter that is at least as small as the pericenter of the planetary orbit}.
Consequently, the moons will naturally be placed onto orbits that bring them close to the WD, causing them to become ideal candidates for future tidal disruption and subsequent pollution of the WD.
We note that the liberated exomoon and parent planets must occupy crossing-orbits, hence future pericenter-passes by the planet will likely cause significant perturbation to the liberated exomoon's orbit, potentially scattering it into/through the Roche surface of the WD.


\section{Discussion}\label{SECN:DISC}

\subsection{Fate of Liberated Moons}\label{SECN:DISC:FATE} 
We have demonstrated that exomoons will be liberated from planetary orbits, both during close-planet-planet encounters and at close-pericenter passage between the planet and WD.
The fate of these liberated moons remains an open question, as they may, 
(i) remain in orbit about the WD; 
(ii) be scattered out of the system by a planet;
(iii) collide with other moons or planets, fragmenting and adding to the debris already in the system, or
(iv) be scattered within the Roche radius of the WD.

If (i) occurs, and an exomoon remains in orbit about the WD, the exomoon may still contribute to pollution of the WD by acting as a perturbative agent on other smaller bodies in the system, going on to (e.g.) scatter members of a planetesimal belt onto WD-crossing orbits in a manner similar to the planet-planetesimal pollution of WDs studied by Bonsor et al. 2011.
We note that such perturbations from exomoons may be particularly efficient, as the exomoons are likely to be scattered by their parent planets onto rather eccentric orbits which may initially be in highly non-equilibrium configurations with respect to such a population of small bodies. 

If (iii) ultimately occurs, and collisional debris is created, then we note that as the mass of moons in the Solar System is $\sim 1,000\times$more than two orders of magnitude greater than the mass of objects in the asteroid belt, the amount of debris created in the system will be large. 
Moreover, this large mass of debris will initially occupy unstable (planet-crossing) orbits and will itself be scattered by planets, with some fraction possibly being scattered towards the WD.

If (iv) ultimately occurs, and the entire mass of a moon enters the Roche radius of the WD, then we stress again, that a large mass exomoon provides a huge reservoir of material to pollute the WD. 
We note that it is far more likely for an exomoon to encounter the Roche surface of an WD than the physical surface of a WD.
Hence it is most likely that in this scenario an exomoon would be tidally disrupted (similar to the asteroids studied in \citet{veretal2014c} and \citet{veretal2015b}), perhaps resulting in an observational signiature qualitatively similar to that seen in the disrupting object(s) around WD 1145+017 (Vanderberg et al. 2015).
A liberated moon which collides with the WD would likely produce an observable transient \citep{beasok2013,disetal2015} as  well as extreme levels of photospheric pollution.  
In DBZ WDs, the mass of this moon would be retained in the convective layer for up to about a Myr, providing an easily observable signature.

We note that, as discussed in Section \ref{SECN:PERI}, exomoons which are released at close-pericenter passage between the planet and WD will naturally have orbits whose pericenter is very close to the Roche radius of the WD.
Subsequent pericenter-passes by the planet will perturb the (now WD-centered) orbit of the liberated moon.
Some such perturbations would completely liberate moons, while others would cause it to hit the WD Roche surface and be disrupted. 

It is clear that the unbinding of exomoons from their parent planets will initiate a phase of dynamical evolution which is likely to be strongly \emph{chaotic}. 
The unbound exomoons will initially occupy orbits which cross those of their parent planets, essentially guaranteeing strong scattering events will occur during the subsequent evolution of the system.
The exomoons will effectively be test-particles compared to their parent planets, and are likely to be scattered into orbits that are both highly eccentric and inclined relative to the initial plane of planetary orbits.
However, compared to other typical \emph{small} bodies in planetary systems (asteroids, comets, etc), liberated exomoons are likely to be significantly more massive, contributing significantly to the ``mass-budget'' of available sub-planet mass objects in such systems, as well as being able to perturb the other small bodies.

The specific fate of the exomoons liberated in Sections (4) and (5) will depend on the subsequent details of the dynamical scattering experienced among the various planets, liberated moons, and collisional debris in the system.
Moreover, the \emph{order} of events is important: e.g. exomoons cannot be liberated during close-pericenter passes with the WD if they have already been liberated during previous planet-planet encounters at large distances from the WD. 

In this present proof-of-principle study, we cannot definitively quantify the fraction of liberated exmoons which will go on to eventually occupy WD-grazing orbits and hence we cannot quantify the number which will pollute the atmosphere of the WD. 
In follow-up work we intend to elucidate this issue by following the detailed long-term evolution of the liberated moons as they scatter throughout the planetary system.

\subsection{The Population of Moons}\label{SECN:DISC:POPN} 

In addition to the above discussion of the fate of liberated exomoons, we must also mention the uncertainties regarding how many exomoons will exist, their size distribution and their orbital distribution. 
It is clear that an understanding of these quantities will be essential if we are to go on to understand what fraction of such objects might be liberated by the mechanisms studied in Sections 4 and 5.

While hopes are high that the first exomoon observation will soon occur (e.g. \citet{kipping2015} and references therein), all of these quantities are at present observationally unknown for exomoons and must be extrapolated from the limited knowledge available to us from the Solar System. 

One could ultimately conceive that the pollution of WDs may proceed by multiple paths, with the overall fraction of WDs which are polluted, $f_{\rm WD,Polluted}$, being composed of contributions due to the exomoon mechanism studied here, $f_{\rm WD,Polluted,Exomoon}$, as well as any-and-all other mechanisms discussed in the introduction, $f_{\rm WD,Polluted,Other}$.
Hence we can write
\begin{equation}
f_{\rm WD,Polluted} = f_{\rm WD,Polluted,Exomoon} + f_{\rm WD,Polluted,Other},
\end{equation}
and then decompose the exomoon component into
\begin{equation}
f_{\rm WD,Polluted,Exomoon} = f_{\rm WD,MP} \times f_{\rm Scatter} \times f_{\rm Liberate} \times f_{\rm Pollute}
\end{equation}
where
$f_{\rm WD,MP}$ is the fraction of WDs with multi-planet systems, 
$f_{\rm Scatter}$ is the fraction of those systems which experience planet-planet scattering, 
$f_{\rm Liberate}$ is the fraction of those scattering systems which liberate exomoons, and
$f_{\rm Pollute}$ is the fraction of liberated exomoons which go on to pollute the WD. 

This study effectively demonstrates that $f_{\rm Liberate}$ can be non-zero.
Detailed knowledge of the population of exomoons and/or assumed forms for their orbital distribution would have to be assumed to provide more detailed refinements of $f_{\rm Liberate}$: we defer elaboration to a future investigation. 
Our proposed study mentioned at the end of Section \ref{SECN:DISC:FATE} (to follow the detailed fate of liberated exomoons) will effectively quantify $f_{\rm Pollute}$. 

We note that observations of WD pollution effectively set $f_{\rm WD,Polluted} \approx 0.3$.
We further note that $f_{\rm WD,MP}$ and $f_{\rm Scatter}$ are completely unknown at this point.
We emphasize that $f_{\rm Scatter}$ is \emph{not} the same as the fraction of simulations from (e.g.) \citet{vergae2015} which scatter during the WD phase: such simulations were initialized with conditions which may be far from those present around real WDs

\subsection{Additional Considerations}\label{SECN:DISC:ADDN} 

Gravity may not be the only perturbative force on exomoons.
Although GB mass loss does not alter a moon's orbit with respect to its parent planet, intense GB radiation could affect its motion.  
Despite shadowing effects \citep[see, e.g.][]{rubincam2013}, a small moon (100m-10km) may be spun-up to fission \citep{veretal2014b}.  
In this case, the exomoon would become an exo-ring, and this exo-ring would be subject to a similar type of disruption from close encounters with other planets during gravitational scattering on the WD phase.
If the moon survives spin-up, then GB radiation could alter its orbit
\citep{veretal2015a}, potentially allowing it to drift closer to the edge of the planet's Hill radius.  
Just how the orbital parameters of the moon would be affected by this radiation is nontrivial and requires future exploration.

Included under the umbrella term {\it moon} are (i) bodies which orbit entities smaller than planets, and (ii) double planets (two planets orbiting each other).
In the Solar system, centaurs, Main Belt asteroids, Jupiter trojans, and trans-Neptunian objects have all been observed to contain moons.  
Also, although not yet observed,  double planets may form through close encounters \citep{ochetal2014,lewetal2015}. Further, moons of moons of planets may be formed through a similar mechanism.  Although these more exotic types of moons may contribute negligibly to the total system mass, their existence emphasizes the need to consider different families of bodies in order to determine orbital architectures and mass reservoirs in WD systems.


Finally, the disrupting object(s) around WD 1145+017 \citep{vanetal2015} may themselves be liberated exomoons.  
Further dynamical studies are needed to determine if the provenance of that minor planet lay in a post-MS exo-Kuiper belt, or around a planet.

\section{Conclusions}\label{SECN:CONC}

Questions remain about the dynamical processes which cause WDs to be polluted by remnant planetary material.  
A potentially major source of extant planetary mass in these systems is exomoons.  
We demonstrated in Section 3 that the onset of dynamical instability in post-MS planetary systems causes planets to experience multiple extreme close-approaches within a tiny fraction of their Hill radii. 
We went on to show in Section 4 that exomoons which survive GB evolution can be easily liberated from their parent planets due to this gravitational scattering.  
This result holds despite moons becoming more entrenched inside their parent planet's Hill radius due to post-MS evolution (equation \ref{eqnrh}).  
Furthermore, in Section 5 we showed that moons may also be released directly onto WD-grazing orbits due to the Hill-sphere contraction experienced by highly scattered planets with close-pericenter approaches to the WD.
The liberation of exomoons provides another population of objects which may themselves be thrust into the WD or act as dynamical perturbers for smaller pollutants.


\section*{Acknowledgements}

The authors thank the referee for a timely and thorough report on our manuscript.

All authors gratefully acknowledge the Royal Society, whose funding (grant number IE140641) 
supported the research leading to these results.  
MJP also acknowledges 
Smithsonian 2015 CGPS/Pell Grant, 
NASA Origins of Solar Systems Program grant NNX13A124G, 
NASA Origins of Solar Systems Program grant NNX10AH40G via sub-award agreement 1312645088477, 
and 
BSF Grant Number 2012384.
DV and BTG also benefited by support from the European Union through ERC grant number 320964.  
DV further acknowledges support from the Institute for Theory and Computation at the Harvard-Smithsonian Center for Astrophysics.



\appendix

\section{Definition of Variables} 
\label{APP:DEFN}

For the convenience of the reader, we provide in Table \ref{TAB:DEFN} a list of all of the quantities used throughout the paper. 
\begin{table*}
 \centering
 \begin{minipage}{180mm}
  \caption{Variables used in this paper}
  \label{TAB:DEFN}
  \begin{tabular}{@{}ll@{}}
  \hline
   Variable & Explanation \\
 \hline
 \texttt{A,B,C,D}   & Labels for the 4 sets of close-encounter simulations. \\
 $a_{\rm p}$        & Semi-major axis of parent planet \\
 $a_{\rm m}$        & Semi-major axis of moon's orbit about parent planet \\
 $b$                & Impact parameter between two planets during the planet-planet hyperbolic encounter orbit \\
 $b_{\rm m}$        & Impact parameter between moon and fly-by planet during the planet-planet hyperbolic encounter orbit \\
 $\beta$            & $=V_{\rm m,c}\eta^{0.5}$m, constant of proportionality. \\
 $C$                & Constant of order unity used in defining $r_{\rm R}$ \\
 $\Delta \vec{V}$  & Kick velocity imparted to the satellite during the planet-planet hyperbolic encounter orbit \\
 $\Delta \vec{V}_{\rm m,f} $ &  Kick velocity perturbation imparted to the moon by fly-by planet during the planet-planet hyperbolic encounter orbit \\
 $\Delta \vec{V}_{\rm p,f} $ &  Kick velocity perturbation imparted to the parent planet by fly-by planet during the planet-planet hyperbolic encounter orbit \\
 $e_{\rm p}$        & Eccentricity of parent planet \\
 $e_{\rm m}$        & Eccentricity of moon's orbit about parent planet \\
 $\eta$             & $=\frac{a_{\rm m}}{r_{\rm H}}$, moon's semi-major axis as a fraction of the Hill sphere.\\
 $\eta_{\substack{\rm Min, \\ \rm Imp}}$ & The minimum value of $\eta$ required to be in the impulsive regime (see Equation \ref{EQN:ETA:IMPULSIZE}). \\
 $\eta_{\substack{\rm Min, \\ \rm Eject}}$     &   The minimum value of $\eta$ required for guaranteed ejection in the impulsive regime (see Equation \ref{EQN:ETA:EJECT}).\\
 $f$            & True anomaly of orbit \\
 $g$            & $\equiv 1 + 3 \left(\frac{q}{a_{\rm m}}\right)^2 \left(\frac{V_q}{V_{\rm m,c}}\right)^4$, a useful dynamical quantity  \\
 $G$            & Gravitational constant \\
 $i_{\rm m}$    & Inclination of moon's orbit about parent planet \\
 $k$            & $=q_{\rm, m} / q$, ratio of close-approach distances during planet-planet hyperbolic encounter \\
 $K$            & Fraction of Hill Sphere outside of which moons become unstable \\
 $M_{\rm \oplus}$& Mass of the Earth  \\
 $M_{\rm J}$& Mass of Jupiter  \\
 $M_{\rm \odot}$& Mass of the Sun  \\
 $M_{\rm \star}$& Mass of star  \\
 $M_{\rm \star}^{\rm MS}$ & Mass of star on main sequence \\
 $M_{\rm \star}^{\rm WD}$ & Mass of star after turning into a white dwarf \\
 $M_{\rm p}$    & Mass of parent planet to moon \\
 $M_{\rm f}$    & Mass of fly-by planet  \\
 $M_{\rm m}$    & Mass of moon \\
 $M_{\rm {wind}}$ & Mass of gas from mass-loss wind from WD enclose within orbit of moon.\\
 $n_{\rm m}$    & Number of moons in close-encounter simulations\\ 
 $\mu$          & $=G\left(M_{\rm p}+ M_{\rm f}\right)$   \\
 $\omega_{\rm m}$ & Longitude of pericenter of moon's orbit about parent planet \\
 $\Omega_{\rm m}$ & Longitude of ascending node of moon's orbit about parent planet \\
 $\texttt{M}_{\rm m}$ & Mean longitude of moon's orbit about parent planet \\
 $P_{\rm p}$        & Orbital period of parent planet \\
 $P_{\rm m}$        & Period of moon's orbit about parent planet \\
 $q$ & Pericenter of the planet-planet hyperbolic encounter orbit \\
 $q_{\rm m}$ & Pericenter of the planet-moon hyperbolic encounter orbit \\
 $r_{\rm m}$ & Moon-planet separation \\
 $r_{\rm H}$ & Hill radius of parent planet \\
 $r_{\rm H}^{\rm MS}$ & Hill radius of parent planet when star is on MS\\
 $r_{\rm H}^{\rm WD}$ & Hill radius of parent planet after star becomes a WD\\
 $r_{\rm R}$ & Roche radius of parent planet \\
 $R_{\rm p}$ & Physical radius of parent planet \\
 $\rho_{\rm p}$ & Density of parent planet \\
 $\rho_{\rm m}$ & Density of moon \\
 $\rho_{\rm {wind}}$ & Density of mass-loss wind from WD \\
 $v_{\rm enc}$ & Volume enclosed by moon's orbit about parent planet  \\
 $V_{\rm p,K}^{WD}$ & Circular velocity of planet in orbit around WD \\
 $V_{\rm q}$ & Velocity-at-pericenter of the planet-planet hyperbolic encounter orbit \\
 $V_{\rm \infty}$ & Velocity-at-infinity of the planet-planet hyperbolic encounter orbit \\
 $V_{\rm m,c}$ & Circular velocity of moon's orbit about parent planet  \\
 $V_{\rm wind}$ & Velocity of wind ejected from WD  \\
\hline
\end{tabular}
\end{minipage}
\end{table*}

\section{Enclosed Mass Within Moon Orbit} 
\label{APP:WIND}

In Section \ref{SECN:ML} we discussed the effects on the stellar mass-loss wind on the orbits of moons. 
We here demonstrate that the mass of wind-driven material within the orbit of the moon (about the planet) is negligible compared to the mass of the planet, and hence the orbit of the moon will be negligibly perturbed.

The enclosed mass within the orbit is $M_{\rm wind} = v_{\rm enc} \rho_{\rm wind}$, where $v_{\rm enc}$ is the enclosed volume. 
Due to stability considerations, $V_{\rm enc}$ is maximized when $r_{m} \approx r_{\rm H}/2$ and the orbit is circular.  
Consequently, max$(M_{\rm wind}$) = $\pi r_{\rm H}^3 {\rm max}(\rho_{\rm wind})/6$.  To compute max($\rho_{\rm wind}$), consider that the maximum mass loss rate for any star that becomes a WD is on the order of $10^{-4} M_{\odot}$/yr (see Fig. 2 of \citealt*{musetal2014}).  
By adopting this mass loss rate, we can derive max($\rho_{\rm wind}$) by assuming a spherically symmetric wind and using equation 5 from \cite{donetal2010} or equation 54 from \cite{veretal2015a}.  
Consequently, we find 
max$(\rho_{\rm wind}) = {\rm max}(\dot{M_{\star}})/(4\pi a_{\rm P}^2 v_{\rm wind}) = 3.1\times10^{-45} \left(\frac{a_{\rm p}}{30 {\rm au}}\right)^{-2} M_{\odot}\,m^{-3}$,
assuming that $v_{\rm wind}$ corresponds to the escape speed from a typical WD ($4 \times 10^3$ km/s).   
The maximum enclosed additional mass due to the wind is thus 
max($M_{\rm wind}) \sim 5\times 10^{-8} \left(\frac{a_{\rm p}}{30{\rm au}}\right) M_{\rm P}$. 
Hence the additional enclosed mass due to the wind is negligible compared to the planetary mass.

\section{Analytic Approximations}
\label{SECN:MOONS:ANALYTIC}
We now present some analytic approximations to shed light on the numerical simulations of Section \ref{SECN:MOONS}.

We denote the moon's post-encounter orbital parameters with primes.  The moon may be destroyed by the planet, or escape the planet's grasp (dissociate), if

\begin{eqnarray}
a_{\rm m}(1-e_{\rm m}) &\le& r_{\rm R}
,
\label{esccon1}
 \\
a_{\rm m}(1+e_{\rm m}) &\ge& Kr_{\rm H}^{\rm WD}
\label{esccon2}
\end{eqnarray}

\noindent{}where $K \approx 1/2$ for coplanar prograde satellites, and $K \approx 1$ for coplanar retrograde satellites. 
In general, computing $a_{\rm m}$ and $e_{\rm m}$ is nontrivial.  
In some cases, however, one might be able to utilize the impulse approximation.  
This approximation holds if both the perturber is quick and the encounter timescale is shorter than the moon's orbital period about the planet \citep{zaktre2004,vermoe2012,jacetal2014}.  
This condition is
\begin{eqnarray}
\left(\frac{V_{q}}{V_{m,c}}\right)
&>& 
\frac{1}{2\pi}\left(\frac{q}{a_{\rm m}}\right), \nonumber
\\
\textrm{or,}&& \label{condImp}
\\
a_{\rm m}^{3/2} 
&>& 
\frac
{b\sqrt{G\left(M_{\rm p} + M_{\rm m}\right)} }
{2\pi V_{q} }
\approx
\frac{\sqrt{G M_{\rm p}}}{2\pi}
\left( \frac{q}{V_{q}} \right).
\nonumber
\end{eqnarray}

\noindent{}where the moon's circular velocity $V_{m, {\rm c}} \equiv \sqrt{G(M_{p}+M_{\rm m})/a_{\rm m}}$, and the last term in parenthesis, representing the pericentre timescale, is determined from the close encounter data. 

Now we evaluate the maximum possible value of this important ratio for which the impulse approximation can occur\footnote{The impulse approximation is
valid for arbitrarily small values of $q/V_q$.  
In fact, the impulse approximation is still valid when the perturbing planet flies inside of the moon's orbit.}.  
We compute this quantity at both the Roche radius of the planet (denoted by the subscript ``inner") and the distance beyond which the moon may escape (denoted by the subscript ``outer").
Equations (\ref{esccon1}) and (\ref{esccon2}) imply that moons can occupy orbits ranging over 
$\frac{r_{\rm R}}{1-e_{\rm m}} < a_{\rm m} < \frac{ Kr_{\rm H}^{\rm WD} }{1+e_{\rm m}}$, 
corresponding to a wide-range of timescales. 
Consequently, 
\begin{equation}
{\rm max}\left( \frac{q}{V_q} \right)_{\rm inner}
=
\left(
\frac{2\pi}{\sqrt{GM_{\rm p}}}
\right)
\left(
\frac
{r_{\rm R}}
{1 - e_{\rm m}}
\right)^{3/2},
\end{equation}

\noindent{}and

\[
{\rm max}\left( \frac{q}{V_q} \right)_{\rm outer}
=
\left(
\frac{2\pi}{\sqrt{GM_{\rm p}}}
\right)
\left(
\frac
{Kr_{\rm H}^{\rm WD}}
{1 + e_{\rm m}}
\right)^{3/2}
\]


\begin{equation}
\ \ \ \ \ \ \ \ \ \ \ \ \ \ \ \ \ \ \ \ \,
=
\frac{a_{\rm p}^{\rm WD}}{v_{\rm p,K}^{\rm WD}}
\left(
\frac{2\pi}{\sqrt{3}}
\right)
\left(
\frac
{K\left(1 - e_{\rm p}^{\rm MS}\right)}
{1 + e_{\rm m}}
\right)^{3/2}
,
\end{equation}


\noindent{}which is \emph{not} a function of planetary mass.

For Earth- and Jupiter-mass planets, $r_{\rm R}\sim\,5\times10^{-5}$ au and $\sim\,5\times10^{-4}$ au respectively.
For typical values of $a_{\rm p}^{\rm WD} = 30$ au, and $e_{\rm p} = e_{\rm m} = 0$ we find that 
\begin{equation}
{\rm max}\left( \frac{q}{V_q} \right)_{\substack{\rm inner, \\ \rm Earth-mass}}  \ \, \sim 6\times10^{3}{\rm s}\,\sim 2 \ {\rm hrs}
,
\end{equation}
\begin{equation}
{\rm max}\left( \frac{q}{V_q} \right)_{\substack{\rm inner, \\ \rm Jupiter-mass}} \sim 1.2\times10^{4}{\rm s}\,\sim 4 \ {\rm hrs}
,
\end{equation}
\begin{equation}
{\rm max}\left( \frac{q}{V_q} \right)_{\rm outer} \ \ \ \ \ \ \ \ \, \,  \sim 3\times10^{9}{\rm s}\,\sim 90 \ {\rm yrs}
.
\end{equation}

If, as an example, we consider the regular moon Callisto, with a semi-major axis $\sim 3\times10^{-2}r_{\rm H,J}$, then using Equation \ref{condImp}, any encounter satisfying $\left(\frac{q}{V_q}\right) < \left(3\times10^{-2}\right)^{3/2}\times 90 \ {\rm yr}\,\sim\,170$ days will be impulsive.
This critical value will be reduced to about 50 days after the Sun becomes a WD\footnote{Jupiter will survive the Sun's post-MS
evolution and expand its orbit adiabatically \citep{dunlis1998,verwya2012}; Callisto's orbit will remain unchanged but will be
further inside Jupiter's (new) Hill sphere.}.

If the impulse approximation can be used, then we can relate the primed and unprimed variables with the formalism of \cite{jacetal2014}.  In particular, we are interested in the semimajor axis and eccentricity changes due to impulses.  Based on their formulae, we find that Equation (\ref{esccon2}) is guaranteed to hold, and escape will occur, if the kick speed $| \Delta \vec{V} |$ on the satellite exceeds

\begin{equation}
\frac{| \Delta \vec{V} |}{V_{m, {\rm c}}}
=
1
+
\sqrt
{
\frac{2}{1 + \left( \frac{K r_{\rm H}}{a_{\rm p}} \right)^{-1} }
}
\approx 
1
.
\label{vcrit}
\end{equation}

\noindent{}where the moon's circular velocity $V_{m, {\rm c}} \equiv \sqrt{G(M_{p}+M_{\rm m})/a_{\rm m}}$ and the approximation results from typical values of $Kr_{\rm H}/a_{\rm p}$ being at most on the order of $10^{-2}$.  
Note that as the value in parenthesis approaches infinity, which mirrors the case of $e_{\rm m} \rightarrow 1$, we recover the formulae in \cite{jacetal2014} and \cite{veretal2014a}.  

We derived Equation (\ref{vcrit}) by considering the extreme cases of equations 4, 6 and 8 in \cite{jacetal2014}.  
We set $e_{\rm m} = 0$ and considered kick directions that maximized the magnitude of the kick needed to eject an orbit.

Now we must relate $\Delta \vec{V}$ to the simulation output from \cite{vergae2015}, as well as assumptions about the moon's location.  We proceed by appealing to Rutherford scattering.  First consider that $\Delta \vec{V}$ is a combination of perturbations from a flyby planet on both the moon-hosting planet and the moon itself.  Denote the resulting kick velocities as $\Delta \vec{V}_{\rm mf}$ and $\Delta \vec{V}_{\rm pf}$.  Consequently, when the flyby star is much closer to the moon, $\Delta \vec{V} \approx \Delta \vec{V}_{\rm mf}$.  Alternatively, when the flyby star is much closer to the moon-hosting planet, $\Delta \vec{V} \approx \Delta \vec{V}_{\rm pf}$.  We focus on these two cases only, and further assume that the moon is on an initially circular orbit.

We derive the kick speed in both cases from equations from Pg. 422 of \cite{bintre1987}, as

\[
\left| \Delta \vec{V}_{\rm pf} \right|
=
\frac{
2 G M_{\rm f} V_{\infty} 
}
{
\sqrt{b^2 V_{\infty}^4 + G^2 \left( M_{\rm f} + M_{\rm p} \right)^2}
}
,
\]

\noindent{}or

\[
\frac{\left| \Delta \vec{V}_{\rm pf} \right|}{V_{m,{\rm c}}}
=
\frac{2M_{\rm f}}{M_{\rm p} + M_{\rm m}}
\left(
\frac{V_{\infty}}{V_{m,{\rm c}}}
\right)
\]

\begin{equation}
\ \ \ \ \ \ \ \ \ \ \
\times \left[
\left(
\frac{ M_{\rm f} + M_{\rm p}  }
{ M_{\rm p} + M_{\rm m}  }
\right)^2
+
\left(
\frac{b}{a_{\rm m}}
\right)^2
\left(
\frac{V_{\infty}}{V_{m,{\rm c}}}
\right)^4
\right]^{-1/2}
\end{equation}

\begin{equation}
\ \ \ \ \ \ \ \ \ \ \
\approx
2
\left(
\frac{V_q}{V_{m,{\rm c}}}
\right)
\left[
4
+
\left(
\frac{q}{a_{\rm m}}
\right)^2
\left(
\frac{V_q}{V_{m,{\rm c}}}
\right)^4
\right]^{-1/2}
\label{vpfapprox}
\end{equation}

\noindent{}

\noindent{}and

\[
\left| \Delta \vec{V}_{\rm mf} \right|
=
\frac{
2 G M_{\rm f} V_{\infty} 
}
{
\sqrt{b_{\rm m}^2 V_{\infty}^4 + G^2 \left( M_{\rm f} + M_{\rm m} \right)^2}
}
,
\]

\noindent{}or

\[
\frac{\left| \Delta \vec{V}_{\rm mf} \right|}{V_{m,{\rm c}}}
=
\frac{2M_{\rm f}}{M_{\rm p} + M_{\rm m}}
\left(
\frac{V_{\infty}}{V_{m,{\rm c}}}
\right)
\]

\begin{equation}
\ \ \ \ \ \ \ \ \ \ \
\times \left[
\left(
\frac{ M_{\rm f} + M_{\rm m}  }
{ M_{\rm p} + M_{\rm m}  }
\right)^2
+
\left(
\frac{b_{\rm m}}{a_{\rm m}}
\right)^2
\left(
\frac{V_{\infty}}{V_{m,{\rm c}}}
\right)^4
\right]^{-1/2}
\end{equation}

\begin{equation}
\ \ \ \ \ \ \ \ \ \ \
\approx
2
\left(
\frac{V_q}{V_{m,{\rm c}}}
\right)
\left[
1
+
\left(
\frac{q_{\rm m}}{a_{\rm m}}
\right)^2
\left(
\frac{V_q}{V_{m,{\rm c}}}
\right)^4
\right]^{-1/2}
\label{vmfapprox}
\end{equation}

\noindent{}where $b_{\rm m}$ is the impact parameter of the flyby and moon, 
$q_{\rm m}$ is the pericentre distance of the flyby and moon,
and the 
approximations in equations (\ref{vpfapprox}) and (\ref{vmfapprox}) assume 
$M_{\rm f} \approx M_{\rm p}$,  $M_{\rm m} \ll M_{\rm p}$ and the relations
from equations (\ref{beqq})-(\ref{infeqq}).  The value of $q_{\rm m}$ is
dependent not only on the semimajor axis of the moon, but its location in its
orbit around the planet.

\begin{figure}
\centering
\begin{tabular}{c}
\includegraphics[trim = 0mm 0mm 0mm 0mm, clip, angle=0, width=\columnwidth]{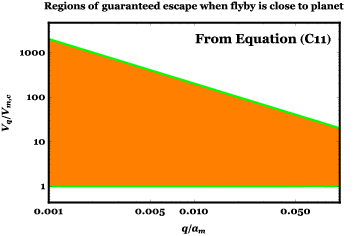}\\
\centerline{}\\
\includegraphics[trim = 0mm 0mm 0mm 0mm, clip, angle=0, width=\columnwidth]{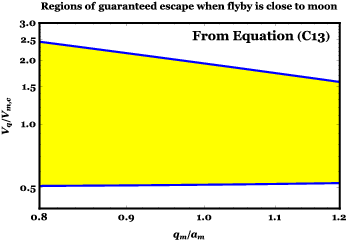}\\
\end{tabular}
\caption{Phase space locations guaranteeing moon escape when the impulse approximation holds (equation \ref{condImp}), where we assume that the mass of the flyby planet is equal to the moon-hosting planet, the moon mass is about $10^{-3}-10^{-2}$ that of the mass of its parent planet, and the dominant kick imparted to the moon-planet system is on the planet (\textit{top panel}; equation \ref{vpfapprox}) or moon (\textit{bottom panel}; equation \ref{vmfapprox}). Different assumptions about the moon mass would shift the $x$-axes on the plots to stay in the appropriate regions of applicability.  These plots suggest that moon escape should be a common occurrence if planet-planet close encounters occur within $0.5r_{\rm H}$ and the pericenter speed between both stars is at least comparable to the circular speed of the moon.
}
\label{EscapeFigs}
\end{figure}

The form of equations (\ref{vpfapprox}) and (\ref{vmfapprox}), which each
are a function of two ratios only, facilitates
comparison with equation (\ref{vcrit}).  Setting equations (\ref{vpfapprox}) 
and (\ref{vmfapprox}) each equal to unity defines surfaces of section
which represents the boundary defining where moon escape is guaranteed to occur.
We plot the guaranteed escape regions in Fig. \ref{EscapeFigs}, and remind the reader that these plots are applicable only in the impulse approximation, where $a_{\rm m}$ is large enough that equation (\ref{condImp}) is satisfied.
Moreover, the top plot of Fig. \ref{EscapeFigs} is applicable when $\left| \Delta \vec{V}_{\rm pf} \right|    \gg   \left| \Delta \vec{V}_{\rm mf} \right|$ , i.e.
\begin{equation}
\ \ \ \ \ \ \ \ \ \ \
\left(
\frac{q}{a_{\rm m}}
\right)^2
\ll   
\left(
\frac{q_{\rm m}}{a_{\rm m}}
\right)^2
- 
3
\left(
\frac{V_q}{V_{\rm m,c}}
\right)^{-4}
\label{EQN:CLOSE:A}
\end{equation}
\noindent{}while the bottom plot of Fig. \ref{EscapeFigs} is applicable when $\left| \Delta \vec{V}_{\rm pf} \right|    \ll   \left| \Delta \vec{V}_{\rm mf} \right|$ , i.e.
\begin{equation}
\ \ \ \ \ \ \ \ \ \ \
\left(
\frac{q}{a_{\rm m}}
\right)^2
\gg   
\left(
\frac{q_{\rm m}}{a_{\rm m}}
\right)^2
- 
3
\left(
\frac{V_q}{V_{\rm m,c}}
\right)^{-4}
\label{EQN:CLOSE:B}
\end{equation}

The figure demonstrates that escape is guaranteed for particular ranges of $V_q/V_{m,{\rm c}}$. 
Near-collisions between both planets are highly destructive to moons, perturbing them out of the system for a range of $V_q/V_{m,{\rm c}}$ that spans a value greater than $a_{m}/q$.
Near collisions between the flyby planet and moon guarantees escape for a more restricted range of $V_q/V_{m,{\rm c}}$ that spans a factor of a few.  
Given the distribution of close encounters in Fig. \ref{FIG:SCATT_DIST_A}, we should expect that moon escape is a common occurrence.

We will find it convenient to define a variable, $k$, such that $q_m\equiv\,k q$ and we emphasize that $k$ can be greater-than or less-than $1$.
We can then rewrite the limiting cases for Equations \ref{EQN:CLOSE:A} and \ref{EQN:CLOSE:B} as  
 \begin{equation}
\ \ \ \ \ \ \ \ \ \ \
\left| \Delta \vec{V}_{\rm pf} \right|      
\overset{\ll}{\underset{\mathrm{\gg}}{=}}
\left| \Delta \vec{V}_{\rm mf} \right|
\label{EQN:CLOSE:C}
\end{equation}
resulting in 
\begin{equation}
\ \ \ \ \ \ \ \ \ \ \
k^2 
\overset{\ll}{\underset{\mathrm{\gg}}{=}}
1 + 3 \left(\frac{q}{a_{\rm m}}\right)^{-2} \left(\frac{V_q}{V_{\rm m,c}}\right)^{-4}
\label{EQN:CLOSE:D}
\end{equation}
 
An important simplification for Equations (\ref{EQN:CLOSE:A}) - (\ref{EQN:CLOSE:D}) comes from noting that 
$a_{\rm m}^2 V_{\rm m,c}^4 = \left(GM_{\rm p}\right)^2$, i.e. it is \emph{not} a function of $a_{\rm m}$.

In addition, we note that even when the impulse approximation does not hold, liberation of the moon may still occur, 
as the impulse approximation is simply a convenient calculation tool.

\begin{table*}
\begin{minipage}[b]{\textwidth}
\centering
\caption{Useful derived quantities for the simulations listed in Table \ref{TAB:MOONS}.
Definitions can be found in Table \ref{TAB:DEFN}}
%
%
\label{TAB:MOONS2}
\begin{tabular}{l cccccccc  } 
\hline
\hline
 	 &	 $q / V_{\rm q}$ 	 &	 $q^2 V_{\rm q}^4$ 	 &	 $\left(GM_{\rm p}\right)^2$ 	 &	 $g \equiv 1 + 3 \frac{ \left(GM_{\rm p}\right)^2}{ q^2 V_{\rm q}^4 }$ 	 &	 $\left| \Delta \vec{V}_{\rm pf} \right|$ 	 &	 $\eta_{\substack{\rm Min, \\  \rm Imp}}$ 	 &	 $V_{\rm m,c}\eta^{0.5}$ 	 &	 $\eta_{\substack{\rm Min, \\ \rm Eject}}$ 	 	 \\
Simulation 	 &	 $(Day)$ 	 &	 $(AU^6 Day^{-4})$ 	 &	 $(AU^6 Day^{-4})$ 	 &	 $(\#)$ 	 &	 $(V_{\rm q})$ 	 &	 $(\#)$ 	 &	 (au/day) 	 &	 $(\#)$ 	 	 \\
\hline
\hline
$\texttt{A}$     &       $1.0 \times 10^{4}$     &       $1.0 \times 10^{-10}$   &       $9.0 \times 10^{-14}$   &       $1.0           $        &       $6.0 \times 10^{-2}$    &       0.4     &       $3.8 \times 10^{-4}$    &       $4.0 \times 10^{1}$                \\
$\texttt{B}$     &       $1.0 \times 10^{3}$     &       $1.0 \times 10^{-12}$   &       $9.0 \times 10^{-14}$   &       $1.3           $        &       $5.1 \times 10^{-1}$    &       0.09    &       $3.8 \times 10^{-4}$    &       $5.5 \times 10^{-1}$               \\
$\texttt{C}$     &       $3.3 \times 10^{3}$     &       $8.1 \times 10^{-15}$   &       $9.0 \times 10^{-14}$   &       $3.4 \times 10^{1}$     &       $9.9 \times 10^{-1}$    &       0.2     &       $3.8 \times 10^{-4}$    &       $1.6                $              \\
$\texttt{X}$     &       $1.0 \times 10^{3}$     &       $1.0 \times 10^{-12}$   &       $8.0 \times 10^{-19}$   &       $1.0           $        &       $1.8 \times 10^{-3}$    &       0.09    &       $5.4 \times 10^{-5}$    &       $9.1 \times 10^{2}$                \\
$\texttt{Y}$     &       $1.0 \times 10^{2}$     &       $7.3 \times 10^{-18}$   &       $8.0 \times 10^{-19}$   &       $1.3           $        &       $5.5 \times 10^{-1}$    &       0.02    &       $5.4 \times 10^{-5}$    &       $1.1 \times 10^{-1}$              \\
$\texttt{Z}$     &       $3.3 \times 10^{1}$     &       $8.1 \times 10^{-19}$   &       $8.0 \times 10^{-19}$   &       $4.0           $        &       $8.9 \times 10^{-1}$    &       0.01    &       $5.4 \times 10^{-5}$    &       $4.1 \times 10^{-2}$              \\
\hline
\end{tabular}
\end{minipage}
\end{table*}

\subsection{Linking Numerical and Analytical results}

We now link our numerical results (from Section 4) with the analytics from Appendix \ref{SECN:MOONS:ANALYTIC}.  
We do so as follows:
Consider the parameter $\eta$, which represents a fraction of the Hill radius ($a_{\rm m}=\eta r_{\rm H}$).

For a planet at $a_{\rm p}=30\,$au, the period of the planet is $P_{\rm p} \simeq 164\,$yrs.
The orbital period of the moon about the planet will be $P_{\rm m}=\left(\frac{1}{3}\right)^{1/2}\eta^{3/2}P_{\rm p} \sim 3.5\times10^4\eta^{3/2}\,$days.
As such, the moon will be in the impulse regime dictated by Equation \ref{condImp} if 
\begin{equation}
\ \ \ \ \ \ \ \ \ \ \
3.5\times10^4 \eta^{3/2} {\rm days} > \left(\frac{q}{V_{\rm q}}\right)
\nonumber
\end{equation}
\noindent{}or
\begin{equation}
\ \ \ \ \ \ \ \ \ \ \
\eta \gsim 10^{-3} \left[\frac{\left(\frac{q}{V_{\rm q}}\right)}{\rm days}\right]^{2/3}.
\label{EQN:ETA:IMPULSIZE}
\end{equation}
\noindent{}We tabulate the values of $\eta$ which satisfy the impulse approximation for Sets \texttt{A}-\texttt{C} and \texttt{X}-\texttt{Z} in Tables \ref{TAB:MOONS} and \ref{TAB:MOONS2}. 

If we consider the values of $q$ and $r_{\rm H}$ in Tables \ref{TAB:MOONS} and \ref{TAB:MOONS2}, then we see that simulation sets \texttt{A} and \texttt{X} have $q$ significantly exterior to $r_{\rm H}/2$, hence are significantly beyond the moons, while sets \texttt{B} and \texttt{C} have $q\approx\,r_{\rm H}/2$, and hence are right at the edge of the region in which the moon's orbit, while \texttt{Y} and \texttt{Z} have $q \ll r_{\rm H}/2$, penetrating deep into the moons' orbital region.
The geometry of these fly-bys means that 
\begin{equation}
\ \ \ \ \ \ \ \ \ \ \
{\rm Geometry} \Rightarrow  
\begin{cases} 
    k^2 \sim 1,         & \mbox{for \texttt{A}} \\ 
    0 < k^2 \lsim 4,    & \mbox{for \texttt{B,C}} \\ 
    k^2 \sim 1,         & \mbox{for \texttt{X}} \\ 
    0 < k^2 \lsim 36,   & \mbox{for \texttt{Y}} \\
    0 < k^2 \lsim 256,  & \mbox{for \texttt{Z}} \\
    \end{cases},
\nonumber
\end{equation}
\noindent{}where $k=q_{\rm m}/q$ .
We evaluate $g \equiv 1 + 3 \left(\frac{q}{a_{\rm m}}\right)^{-2} \left(\frac{V_q}{V_{\rm m,c}}\right)^{-4}$ from Equation \ref{EQN:CLOSE:D} (see Table \ref{TAB:MOONS} and Table \ref{TAB:MOONS2}), and find that the relative magnitude of the kicks will be given by 
\begin{eqnarray}
k^2 \sim g    &\Rightarrow&   \left| \Delta \vec{V}_{\rm pf} \right|   \sim \left| \Delta \vec{V}_{\rm mf} \right| \,\,\, \mbox{for \texttt{A}} 
\\ 
k^2 \overset{\ll}{=} g    &\Rightarrow&  \left| \Delta \vec{V}_{\rm pf} \right|   \sim \left| \Delta \vec{V}_{\rm mf} \right| \,\,\, \mbox{for \texttt{B}} 
\\ 
k^2 \ll g    &\Rightarrow&   \left| \Delta \vec{V}_{\rm pf} \right|   \sim \left| \Delta \vec{V}_{\rm mf} \right| \,\,\, \mbox{for \texttt{C}} 
\\ 
k^2 \sim g    &\Rightarrow&   \left| \Delta \vec{V}_{\rm pf} \right|   \sim \left| \Delta \vec{V}_{\rm mf} \right| \,\,\, \mbox{for \texttt{X}} 
\\ 
k^2 \overset{\ll}{\underset{\mathrm{\gg}}{=}} g    &\Rightarrow&   \left| \Delta \vec{V}_{\rm pf} \right|   \overset{\ll}{\underset{\mathrm{\gg}}{=}} \left| \Delta \vec{V}_{\rm mf} \right| \,\,\, \mbox{for \texttt{Y, Z}} 
\\ 
\nonumber
\end{eqnarray}
\noindent{}
Hence the dominant perturbation will be of order
\begin{equation}
\ \ \ \ \ \ \ \ \ \ \
\left| \Delta \vec{V} \right|  \sim 
\begin{cases} 
    \sim\left| \Delta \vec{V}_{\rm pf} \right| \sim  \BigO{10^{-1}V_{\rm q}},         & \mbox{for \texttt{A}} \\ 
    \sim\left| \Delta \vec{V}_{\rm mf} \right| \sim \BigO{V_{\rm q}},         & \mbox{for \texttt{B, C}} \\ 
    \sim\left| \Delta \vec{V}_{\rm pf} \right| \sim \BigO{10^{-3}V_{\rm q}},         & \mbox{for \texttt{X}} \\ 
    \sim\max\left( \left| \Delta \vec{V}_{\rm pf} \right|,\left| \Delta \vec{V}_{\rm mf} \right|\right)         \sim \BigO{V_{\rm q}},         & \mbox{for \texttt{Y,Z}} \\ 
    \end{cases}.
\label{OrderDV}
\end{equation}

We wish to understand whether the condition for ejection in the impulsive regime in Equation \ref{vcrit} ($\left| \Delta \vec{V} \right|  \gsim  V_{\rm m,c}$) is satisfied for the cases in Equation \ref{OrderDV}. 
Hence we use the $\eta$-dependant expressions for $V_{\rm m,c}=\beta\eta^{-0.5}$ from Table \ref{TAB:MOONS2} to find that $\left| \Delta \vec{V} \right|  \gsim  V_{\rm m,c}$ implies 
\begin{equation}
\ \ \ \ \ \ \ \ \ \ \
\eta \gsim \left(\frac{\beta}{\left| \Delta \vec{V} \right|}\right)^2
\label{EQN:ETA:EJECT}
\end{equation}
\noindent{}We evaluate Equation \ref{EQN:ETA:EJECT} using the Equation \ref{OrderDV} and tabulate the values of $\eta$ which will satisfy this ejection criterion for simulation sets \texttt{A}-\texttt{C} and \texttt{X}-\texttt{Z} in Table \ref{TAB:MOONS}. 
The tabulated values for $\eta_{\substack{\rm Min, \\ \rm Eject}}$ (the minimum value of $\eta$ required for guaranteed ejection in the impulsive regime) make clear that this can \emph{never} be satisfied for \texttt{A, C} and \texttt{X}, as $\eta_{\substack{\rm Min, \\ \rm Eject}}\geq1.0$.
In contrast, ejection seems almost guaranteed for the majority of moon orbits in \texttt{Z}.
However, we note that \emph{significant} ejection occurs in simulation set \texttt{C}, \emph{despite} the majority of moons not being in the impulsive regime, highlighting the additional insight numerical simulations can provide. 

Comparison of the analytic approximations in this section with the numerical results depicted in Figure \ref{FIG:MOON:SURVIVE}, confirm that close planetary approaches (within the Hill Sphere) will efficiently eject moons.

\label{lastpage}
\end{document}